\documentclass[reprint,aps,prb,superscriptaddress,showpacs]{revtex4-1}
\pdfoutput=1

\usepackage{bm}
\usepackage{amsmath,amsthm,amssymb}
\usepackage{array,booktabs,longtable}
\usepackage{mathrsfs}
\usepackage{subfigure}
\usepackage{color}
\usepackage{tikz}
\usetikzlibrary{decorations.markings}
\usetikzlibrary{calc,shapes,positioning,arrows,snakes}
\usepackage{array}
\usepackage{verbatim}
\usepackage{dsfont}
\usepackage{ifthen}

\newcommand{\bra}[1]{\langle #1 |}
\newcommand{\ket}[1]{| #1\rangle}

\DeclareMathOperator{\tr}{tr}

\newcommand{\bmm}{\begin{matrix}}
\newcommand{\emm}{\end{matrix}}
\makeatletter
\newcommand{\thickhline}{%
    \noalign {\ifnum 0=`}\fi \hrule height 1.25pt
    \futurelet \reserved@a \@xhline
}
\newcolumntype{"}{@{\hskip\tabcolsep\vrule width 1.25pt\hskip\tabcolsep}}
\makeatother

\newcommand{\Oop}[1]{
\begin{tikzpicture}
[scale=0.28,baseline={([yshift=-.5ex]current bounding box.center)}]
\path[draw,thick,gray!70] (-1,-1) rectangle (1,1);
\draw (0,0) node {$#1$};
\begin{scope}[rounded corners=12pt,thick,rotate=50,decoration={
    markings,
    mark=at position 0.5 with {\arrow{}}}
    ] 
    \path[draw,postaction={decorate}] (-2,0) -- (0,2) -- (2,0) -- (0,-2) -- cycle;
    \path[draw,postaction={decorate}] (0,-2) -- (-2,0) -- (0,2) -- (2,0) -- cycle;
    \path[draw,postaction={decorate}] (2,0) -- (0,-2) -- (-2,0) -- (0,2) -- cycle;
    \path[draw,postaction={decorate}] (0,2) -- (2,0) -- (0,-2) -- (-2,0) -- cycle;        
\end{scope}

\end{tikzpicture}
}

\newcommand{\OopNumb}[1]{
\begin{tikzpicture}
[scale=0.28,baseline={([yshift=-.5ex]current bounding box.center)}]
\path[draw,thick,gray!70] (-1,-1) rectangle (1,1);
\draw (0,0) node {$#1$};
\draw (-1.3,-1.3) node {\scalebox{0.8}{$4$}};
\draw (-1.3,1.3) node {\scalebox{0.8}{$1$}};
\draw (1.3,-1.3) node {\scalebox{0.8}{$3$}};
\draw (1.3,1.3) node {\scalebox{0.8}{$2$}};

\begin{scope}[rounded corners=12pt,thick,rotate=50,decoration={
    markings,
    mark=at position 0.5 with {\arrow{}}
  }
    ] 
    \path[draw,postaction={decorate}] (-2,0) -- (0,2) -- (2,0) -- (0,-2) -- cycle;
    \path[draw,postaction={decorate}] (0,-2) -- (-2,0) -- (0,2) -- (2,0) -- cycle;
    \path[draw,postaction={decorate}] (2,0) -- (0,-2) -- (-2,0) -- (0,2) -- cycle;
    \path[draw,postaction={decorate}] (0,2) -- (2,0) -- (0,-2) -- (-2,0) -- cycle;        
\end{scope}

\end{tikzpicture}
}

\newcommand{\ZredLine}[1]{
\begin{tikzpicture}
[scale=0.28,baseline={([yshift=-.5ex]current bounding box.center)},rounded corners=12pt,red,thick,decoration={
    markings,
    mark=at position 0.5 with {\arrow{}}}]
    \path[draw,postaction={decorate}] (0,0) -- (2,2);
    \draw (1+0.05,1-0.2) node[right]{\scalebox{0.8}{$#1$}};
    \draw[fill=red] (1,1) circle (0.2cm);
\end{tikzpicture}
}
\newcommand{\ZblueLine}[1]{
\begin{tikzpicture}
[scale=0.28,baseline={([yshift=-.5ex]current bounding box.center)},rounded corners=12pt,blue,thick,decoration={
    markings,
    mark=at position 0.5 with {\arrow{}}}]
    \path[draw,postaction={decorate}] (0,0) -- (2,2);
    \draw (1+0.05,1-0.2) node[right]{\scalebox{0.8}{$#1$}};
    \draw[fill=blue] (1,1) circle (0.2cm);
\end{tikzpicture}
}

\newcommand{\XredLine}[1]{
\begin{tikzpicture}
[scale=0.28,baseline={([yshift=-.5ex]current bounding box.center)},rounded corners=12pt,thick,red,decoration={
    markings,
    mark=at position 0.5 with {\arrow{}}}]
    \path[draw,postaction={decorate}] (0,2) -- (2,0);
    \draw (1+0.05,1+0.2) node[right]{\scalebox{0.8}{$#1$}};
    \draw[fill=red] (1,1) circle (0.2cm);
\end{tikzpicture}
}

\newcommand{\XblueLine}[1]{
\begin{tikzpicture}
[scale=0.28,baseline={([yshift=-.5ex]current bounding box.center)},rounded corners=12pt,thick,blue,decoration={
    markings,
    mark=at position 0.5 with {\arrow{}}}]
    \path[draw,postaction={decorate}] (0,2) -- (2,0);
    \draw (1+0.05,1+0.2) node[right]{\scalebox{0.8}{$#1$}};
    \draw[fill=blue] (1,1) circle (0.2cm);
\end{tikzpicture}
}

\newcommand{\OopRed}[1]{
\begin{tikzpicture}
[scale=0.28,baseline={([yshift=-.5ex]current bounding box.center)}]
\path[draw,thick,gray!70] (-1,-1) rectangle (1,1);
\draw (0,0) node {$#1$};
\begin{scope}[rounded corners=12pt,thick,red,rotate=50,decoration={
    markings,
    mark=at position 0.5 with {\arrow{}}}
    ] 
    \path[draw,postaction={decorate}] (-2,0) -- (0,2) -- (2,0) -- (0,-2) -- cycle;
    \path[draw,postaction={decorate}] (0,-2) -- (-2,0) -- (0,2) -- (2,0) -- cycle;
    \path[draw,postaction={decorate}] (2,0) -- (0,-2) -- (-2,0) -- (0,2) -- cycle;
    \path[draw,postaction={decorate}] (0,2) -- (2,0) -- (0,-2) -- (-2,0) -- cycle;        
\end{scope}

\end{tikzpicture}
}

\newcommand{\OopBlue}[1]{
\begin{tikzpicture}
[scale=0.28,baseline={([yshift=-.5ex]current bounding box.center)}]
\path[draw,thick,gray!70,fill=gray!20] (-1,-1) rectangle (1,1);
\draw (0,0) node {$#1$};
\begin{scope}[rounded corners=12pt,thick,blue,rotate=50,decoration={
    markings,
    mark=at position 0.5 with {\arrow{}}}
    ] 
    \path[draw,postaction={decorate}] (-2,0) -- (0,2) -- (2,0) -- (0,-2) -- cycle;
    \path[draw,postaction={decorate}] (0,-2) -- (-2,0) -- (0,2) -- (2,0) -- cycle;
    \path[draw,postaction={decorate}] (2,0) -- (0,-2) -- (-2,0) -- (0,2) -- cycle;
    \path[draw,postaction={decorate}] (0,2) -- (2,0) -- (0,-2) -- (-2,0) -- cycle;        
\end{scope}

\end{tikzpicture}
}

\newcommand{\EdgeRed}[1]{
\begin{tikzpicture}
[scale=0.28,baseline={([yshift=-.5ex]current bounding box.center)}]
\path[draw,thick,white] (-1,-1) rectangle (1,1);
\path[draw,thick,gray!70,fill=gray!20] (-3,-1) rectangle (-1,1);
\draw (0,0) node {$#1$};
\draw[very thick,black!80,dashed] (-1,-1.8) -- (-1,1.8);
\begin{scope}[rounded corners=12pt,thick,red,rotate=0,decoration={
    markings,
    mark=at position 0.52 with {\arrow{}}}
    ] 
    \path[draw,postaction={decorate}] (0,-2) -- (-2,0) -- (0,2);       
\end{scope}

\end{tikzpicture}
}

\newcommand{\EdgeBlue}[1]{
\begin{tikzpicture}
[scale=0.28,baseline={([yshift=-.5ex]current bounding box.center)}]
\path[draw,thick,white,fill=gray!20] (-1,-1) rectangle (1,1);
\path[draw,thick,gray!70] (-3,-1) rectangle (-1,1);
\draw (0,0) node {$#1$};
\draw[very thick,black!80,dashed] (-1,-1.8) -- (-1,1.8);
\begin{scope}[rounded corners=12pt,thick,blue,rotate=0,decoration={
    markings,
    mark=at position 0.52 with {\arrow{}}}
    ] 
    \path[draw,postaction={decorate}] (0,-2) -- (-2,0) -- (0,2);       
\end{scope}

\end{tikzpicture}
}

\newcommand{\GammaRed}{
\begin{tikzpicture}
[scale=0.22,baseline={([yshift=-.5ex]current bounding box.center)}]
\path[draw,thick,white,fill=gray!20] (-1.5,-3.5) rectangle (-0.5,-2.5);
\path[draw,thick,white,fill=gray!20] (-1.5,-1.5) rectangle (-0.5,-0.5);
\path[draw,thick,white,fill=gray!20] (-0.5,-2.5) rectangle (0.5,-1.5);
\path[draw,thick,white,fill=gray!20] (-0.5,-0.5) rectangle (0.5,0.5);
\begin{scope}[rounded corners=1.5pt,thick,red,rotate=0,decoration={
    markings,
    mark=at position 0.55 with {\arrow{}}}
    ] 
    \path[draw,postaction={decorate}] (0,-4) -- (-1,-3) -- (0,-2) -- (-1,-1) -- (0,0) -- (-1,1);       
\end{scope}
\end{tikzpicture}
}

\newcommand{\GammaBlue}{
\begin{tikzpicture}
[scale=0.22,baseline={([yshift=-.5ex]current bounding box.center)}]
\path[draw,thick,white,fill=gray!20] (-1.5,-2.5) rectangle (-0.5,-1.5);
\path[draw,thick,white,fill=gray!20] (-1.5,-0.5) rectangle (-0.5,0.5);
\path[draw,thick,white,fill=gray!20] (-0.5,-3.5) rectangle (0.5,-2.5);
\path[draw,thick,white,fill=gray!20] (-0.5,-1.5) rectangle (0.5,-0.5);
\begin{scope}[rounded corners=1.5pt,thick,blue,rotate=0,decoration={
    markings,
    mark=at position 0.55 with {\arrow{}}}
    ] 
    \path[draw,postaction={decorate}] (0,-4) -- (-1,-3) -- (0,-2) -- (-1,-1) -- (0,0) -- (-1,1);       
\end{scope}
\end{tikzpicture}
}

\newcommand{\TwoBlocks}[1]{
\begin{tikzpicture}
[scale=0.5,baseline={([yshift=-.5ex]current bounding box.center)}]
\ifnum#1=1  
        \path[draw,white,fill=gray!20] (0,0) rectangle (0+1,0+1);
        \path[draw,white] (0,1) rectangle (0+1,1+1);
        \draw (.5,.5) node {$\tilde p$};
        \draw (.5,1.5) node {$p$};
        \draw[thick, dotted] (0,-.5) -- (0,2.25);
    \else
        \path[draw,white] (0,0) rectangle (0+1,0+1);
        \path[draw,white,fill=gray!20] (0,1) rectangle (0+1,1+1);
        \draw (.5,.5) node {$p$};
        \draw (.5,1.5) node {$\tilde p$};
        \draw[thick, dotted] (0,-.25) -- (0,2.5);
\fi

\draw[ultra thick] (0,0) -- (0,2);

\end{tikzpicture}
}

\newcommand{\ComEdge}[3]{
\begin{tikzpicture}
[scale=0.22,baseline={([yshift=-.5ex]current bounding box.center)}]
\begin{scope}[rounded corners=12pt,thick,rotate=0,decoration={
    markings,
    mark=at position 0.52 with {\arrow{}}}
    ]
\ifnum#1=1 
    \path[draw,#2,postaction={decorate}] (0,-2) -- (-2,0) -- (0,2);
    \path[draw,white,double=black] (0,-4) -- (-2,-2) -- (0,0);
    \path[draw,#3,postaction={decorate}] (0,-4) -- (-2,-2) -- (0,0);
\else
    \path[draw,#3,postaction={decorate}] (0,-4) -- (-2,-2) -- (0,0);
    \path[draw,white,double=black] (0,-2) -- (-2,0) -- (0,2);
    \path[draw,#2,postaction={decorate}] (0,-2) -- (-2,0) -- (0,2);
\fi
\end{scope}
\end{tikzpicture}
}

\begin{document}

% Use the \preprint command to place your local institutional report
% number in the upper righthand corner of the title page in preprint mode.
% Multiple \preprint commands are allowed.
% Use the 'preprintnumbers' class option to override journal defaults
% to display numbers if necessary
%\preprint{}

%Title of paper
\title{Edge-entanglement spectrum correspondence in a nonchiral topological phase and Kramers-Wannier duality}

% repeat the \author .. \affiliation  etc. as needed
% \email, \thanks, \homepage, \altaffiliation all apply to the current
% author. Explanatory text should go in the []'s, actual e-mail
% address or url should go in the {}'s for \email and \homepage.
% Please use the appropriate macro foreach each type of information

% \affiliation command applies to all authors since the last
% \affiliation command. The \affiliation command should follow the
% other information
% \affiliation can be followed by \email, \homepage, \thanks as well.
%\author{}
%\email[]{Your e-mail address}
%\homepage[]{Your web page}
%\thanks{}
%\altaffiliation{}

\author{Wen Wei Ho}
\affiliation{Perimeter Institute for Theoretical Physics, Waterloo, Ontario, N2L 2Y5 Canada}

\author{Lukasz Cincio}
\affiliation{Perimeter Institute for Theoretical Physics, Waterloo, Ontario, N2L 2Y5 Canada}

\author{Heidar Moradi}
\affiliation{Perimeter Institute for Theoretical Physics, Waterloo, Ontario, N2L 2Y5 Canada}

\author{Davide Gaiotto}
\affiliation{Perimeter Institute for Theoretical Physics, Waterloo, Ontario, N2L 2Y5 Canada}

\author{Guifre Vidal}
\affiliation{Perimeter Institute for Theoretical Physics, Waterloo, Ontario, N2L 2Y5 Canada}

\date{\today}

\begin{abstract}
In a system with chiral topological order, there is a remarkable correspondence between the edge and entanglement spectra: the low-energy spectrum of the system in the presence of a physical edge coincides with the lowest part of the entanglement spectrum (ES) across a virtual cut of the system into two parts, up to rescaling and shifting. This correspondence is believed to be due to the existence of protected gapless edge modes. In this paper, we explore whether the edge-entanglement spectrum correspondence extends to nonchiral topological phases, where there are no protected gapless edge modes. Specifically, we consider the Wen-plaquette model, which is equivalent to the Kitaev toric code model and has $\mathbb{Z}_2$ topological order (quantum double of $\mathbb{Z}_2$). The unperturbed Wen-plaquette model displays an exact correspondence: both the edge and entanglement spectra within each topological sector $a$ ($a = 1,\cdots,4$) are flat and equally degenerate. Here, we show, through a detailed microscopic calculation, that in the presence of generic local perturbations: (i) the effective degrees of freedom for both the physical edge and the entanglement cut consist of a (spin-$1/2$) spin chain, with effective Hamiltonians $H_{\text{edge}}^a$ and $H_{\text{ent.}}^a$, respectively, both of which have a $\mathbb{Z}_2$ symmetry enforced by the bulk topological order; (ii) there is in general no match between the low energy spectra of $H_{\text{edge}}^a$ and $H_{\text{ent.}}^a$, that is, there is no edge-ES correspondence. However, if supplement the $\mathbb{Z}_2$ topological order with a global symmetry (translational invariance along the edge/entanglement cut), i.e., by considering the Wen-plaquette model as a symmetry-enriched topological phase (SET), then there is a finite domain in Hamiltonian space in which both $H_{\text{edge}}^a$ and $H_{\text{ent.}}^a$ realize the critical Ising model, whose low-energy effective theory is the $c = 1/2$ Ising CFT. This is achieved because the presence of the global symmetry implies that the effective degrees of freedom of both the edge and entanglement cut are governed by Kramers-Wannier self-dual Hamiltonians, in addition to them being $\mathbb{Z}_2$ symmetric, which is imposed by the topological order. Thus, by considering the Wen-plaquette model as a SET, the topological order in the bulk together with the translation invariance of the perturbations along the edge/cut imply an edge-ES correspondence at least in some finite domain in Hamiltonian space.
\end{abstract}

% insert suggested PACS numbers in braces on next line
% insert suggested keywords - APS authors don't need to do this
%\keywords{}

%\maketitle must follow title, authors, abstract, \pacs, and \keywords
\maketitle

% body of paper here - Use proper section commands
% References should be done using the \cite, \ref, and \label commands

\section{Introduction}
\label{sect:Introduction}
Quantum entanglement has been found to be very useful in characterizing topological states of matter, which is not possible with conventional local order parameters. In particular, the entanglement spectrum (ES)\cite{LiHaldane2008} is one such tool, and has been applied to many systems, such as quantum Hall fluids\cite{LiHaldane2008, ESfqheTorus, RegnaultBernevigHaldaneESJainStates, ThomaleSterndyniakRegnaultBernevigEGap, DubailReadRezayiESRealSpace, RodSimonSlingerlandESRealSpace, SterndyniakChandranEtAlRealSpace, BulkESFQH}, topological order\cite{YaoQiESofKitaevModel, QiKatsuraLudwig2012, HsiehFuES},  topological insulators \cite{FidowskiTopInsScES, TurnerZhangVishwanathTopInsulatorES, ProdanHughesBernevigES}, fractional Chern insulators\cite{RegnaultBernevigFracChernIns}, symmetry-protected topological phases \cite{PollmannBergTurnerOshikawaSPT,RaoWanZhangESSPT}, quantum spin chains\cite{NienhuisCampostriniCalabreseSpinChain, ThomaleArovasBernevigSpinChains, DeChiaraLeporiEtAlSpinChains,TorlaiTagliacozzoDeChiaraESChain} and ladders \cite{PoilblancESLadder, JamesKonikESThroughLadders, LauchliSchliemannESLadder, ChenFradkinESLadders, LundgrenFujiFurukawaESLadders, ES_Tanaka}, and other spin and fermionic systems.
%Li & Haldane
% Many citations in Ludwig... Qi

\subsection{Entanglement spectrum and edge-ES correspondence}
The ES is defined as follows. Given a system's ground state $\ket{\Psi}$, together with a bipartition of the full Hilbert space $\mathcal{H}$ into parts $L$ and $R$, so that $\mathcal{H} = \mathcal{H}_L \otimes \mathcal{H}_R$, one forms the reduced density matrix $\rho_L := \Tr_R \ket{\Psi} \bra{\Psi}$ on $L$. Because of hermiticity and positivity, the reduced density matrix can be written in the thermal form $\rho_L \equiv \frac{1}{Z} e^{-H_{\text{ent.}}}$, where $H_{\text{ent.}}$ is the so-called entanglement Hamiltonian and $Z = \text{Tr}(e^{-H_{\text{ent.}}})$. The ES is then simply the eigenenergies of the entanglement Hamiltonian.

This definition of the ES is an operational one. However, there exists a remarkable observation made by Li and Haldane \cite{LiHaldane2008} for quantum Hall systems and by others in subsequent works for $(2+1)$-d topological phases: in the cases where the system possesses low energy states living near an open boundary of the manifold the system is placed on (i.e. edge states), it was found that the low-lying edge spectrum of the physical boundary Hamiltonian on $L$ are in one-to-one correspondence with the low-lying spectrum of $H_{\text{ent.}}$, a so-called edge-ES correspondence. (This correspondence should not be confused with the more established bulk-edge correspondence \cite{HalperinEdgeIQHE,WenFQHEEdge92,WenTopOrdersEdge1995} also used in the context of topological phases).

Analytic proofs of the edge-ES correspondence have been proposed, for example in \Ref{QiKatsuraLudwig2012} for $(2+1)$-d topological states whose edge states are described by a $(1+1)$-d CFT. In that work, a `cut and glue' approach and methods of boundary CFT were used, and it was claimed that the edge and entanglement spectra should be equal up to rescaling and shifting in the low energy limit. However, this method is only applicable to chiral topological phases, where there are protected, physical chiral edge states appearing at an actual spatial boundary of a system. For the case of a non-chiral topological phase, it is unclear as to what information the ES will yield, or even if there is any form of the edge-ES correspondence that exists\cite{UniversalES}.
% Qi, Katsura, and Ludwig
% Need to check

\subsection{Edge-ES correspondence in non-chiral topological order}
 It is thus the purpose of this paper to explore the edge-ES correspondence in non-chiral topological phases. Specifically we consider the $\mathbb{Z}_2$ Wen-plaquette model\cite{WenPlaquetteModel2003} (unitarily equivalent to Kitaev's toric code model\cite{Kitaev20032} in the bulk), and ask if some form of correspondence exists.  We choose to work on an infinite cylinder with a bipartition into two semi-infinite cylinders terminated with smooth edges. The model on this geometry has four topological sectors $a$ ($a = 1,\cdots,4$) with four locally indistinguishable ground states (these are states with well-defined anyonic flux), and therefore the edge theory on the semi-infinite cylinder and ES of the full cylinder can be unambiguously defined within each topological sector. For the unperturbed Wen-plaquette model, there is in fact an {\it exact} edge-ES correspondence because the edge and entanglement spectra in each $a$-sector are flat and equally degenerate; thus, the two spectra agree perfectly up to rescaling and shifting. However, this correspondence is potentially lost in the presence of perturbations. Here, we present a detailed microscopic derivation of the edge and entanglement Hamiltonians of the Wen-plaquette model deformed by generic local perturbations, which allows us to compare the two spectra and hence explore the edge-ES correspondence. This calculation constitutes the main result of our paper. 

From our calculation, we find the following.

\vspace{5pt}
\noindent {\bf (i)} Our calculation shows that the edge states (belonging to the lowest energy eigenspace) of the unperturbed Wen-plaquette model on the semi-infinite cylinder are generated by so-called boundary operators, which can be mapped to $\mathbb{Z}_2$ symmetric operators acting on a finite length (spin-$1/2$) spin chain, the effective low-energy degrees of freedom. The effects of generic local perturbations to the Wen-plaquette model are to lift this degeneracy - we find that the effective Hamiltonian in each topological sector $a$ acting on these edge states, $H_{\text{edge}}^a$, is a $\mathbb{Z}_2$ symmetric Hamiltonian acting on the spin chain. The $\mathbb{Z}_2$ symmetry can be understood as arising from the bulk topological order: it is generated by a Wilson loop operator wrapping around the cylinder. 

\vspace{5pt}
\noindent {\bf (ii)} We also find that the entanglement Hamiltonian $H_{\text{ent.}}^a$ in each $a$-sector acts on a (spin-$1/2$) spin chain of equal length, and is generated in part by the edge Hamiltonians $( H_{\text{edge},L}^a + H_{\text{edge},R}^a )$ of the two halves of the bipartition ($L$ and $R$) and in part by $V_{LR}$, a perturbation spanning the cut. It is also $\mathbb{Z}_2$ symmetric. However, $H_{\text{ent.}}^a$ is in general not equal to the edge Hamiltonians, being different in some arbitrary way. Thus, there is in general no edge-ES correspondence for generic perturbations, even in the low energy limit, i.e. the low lying values of the edge and entanglement spectra do not match. 

\vspace{5pt}
\noindent {\bf (iii)} We do find a mechanism in which an edge-ES correspondence is established, though. If we consider the Wen-plaquette model as a symmetry enriched topological phase (SET), by supplementing the $\mathbb{Z}_2$ topological order with a global translational symmetry along the edge/entanglement cut, achieved by restricting perturbations to those that respect the symmetry, then there is a finite domain in Hamiltonian space in which both $H_{\text{edge}}^a$ and $H_{\text{ent.}}^a$ realize the critical $(1+1)$-d Ising model, which has the $c = 1/2$ Ising CFT as its low energy effective theory. It is in this context that we have observed, in concrete examples, the 	edge-ES correspondence being realized. This happens because the global translational symmetry implies that the effective degrees of freedom of both the edge and entanglement cut are governed by Kramers-Wannier self-dual Hamiltonians, in addition to them being $\mathbb{Z}_2$ symmetric, which is imposed by the topological order. The fact that the Hamiltonians have $\mathbb{Z}_2$ symmetry and Kramers-Wannier self-duality then further guarantees that all perturbations about the $c = 1/2$ Ising CFT must be irrelevant, giving us the result that the low lying values of the edge and entanglement spectra match upon shifting and rescaling. We therefore see that by considering the Wen-plaqeutte model as a SET, the topological order in the bulk together with the translation invariance of the perturbations along the edge/cut guarantee an edge-ES correspondence at least in some finite domain in Hamiltonian space.

It should be noted that there have been studies of the edge theories and ES of two-dimensional spin systems within the framework of projected entangled pair state models (PEPs)\cite{Shuo, ES_AKLT},  but our paper uses standard techniques in perturbation theory and therefore offers  a complementary approach to probing the edge-ES correspondence.

\subsection{Structure of paper}
The rest of the paper is organized as follows. In Sec.~\ref{sect:Exact}, we introduce the Wen-plaquette model and solve for its edge theory on the semi-infinite cylinder by identifying boundary operators and mapping them to $\mathbb{Z}_2$ symmetric operators acting on a finite length (spin-$1/2$) spin chain. We also calculate the entanglement spectra on the infinite cylinder by deriving an effective spin ladder Hamiltonian whose ground states equal the ground states on the infinite cylinder. Next, in Sec.~\ref{sect:Perturbed}, we consider the effects of perturbations to the Wen plaquette model. We present a quick summary of the Schrieffer-Wolff (SW) transformation, central to the derivation of our results. Then, we derive the edge theory and solve for the entanglement spectrum. This allows us to compare the edge-ES correspondence for the perturbed Wen-plaquette model. Then, in Sec.~\ref{sect:KW}, we identify the mechanism to establish an edge-ES correspondence: we consider the Wen-plaquette model as a symmetry enriched topological phase (SET) with a global translational symmetry along the edge/entanglement cut, which forces the edge and entanglement Hamiltonians to be additionally Kramers-Wannier self-dual, resulting in the edge-ES correspondence. We also provide a numerical example of the correspondence where the perturbations are uniform magnetic fields acting on single spins. Lastly, in Sec.~\ref{sect:Conclusion}, we discuss the implications of our findings and conclude. Appendix \ref{sect:AppendixA} presents the Schrieffer-Wolff transformation and necessary formulas, appendix \ref{sect:AppendixB} presents the perturbation theory calculations for the entanglement Hamiltonian, specifically, $\Lambda'$, defined in Sec.~\ref{sect:pertES}, while Appendix \ref{sect:AppendixC} presents the derivation of the edge and entanglement Hamiltonians of the Wen-plaquette model on an infinite cylinder perturbed by uniform single-site magnetic fields, as considered in Sec.~\ref{sect:KW}.

% Spiel about topological phase, quantum entanglement, what is entaglement spectrum, bulk-edge correspondence
% Differences between chiral and non-chiral topological phases
% Talk about proofs (Qi, Ludwig) (chiral only)
% Coupled ladder systems, cut-and-glue approach

% Paper's thesis: proof of this correspondence.
% Model: Wen-plaquette model, add perturbations
% Main idea: reduce to two coupled Ising chains which are degrees of freedom on boundary.
% Derive effective Hamiltonian (local algebra)
% Perturbation theory; roles of MES: project into the different Z_2 sectors, and a twisted Ising chain?

% wen-plaquette IS the Ising.

% Outline: 
% Hilbert space structure
% Edge Hamiltonian (after perturbations)
% Entanglement Hamiltonian
% Examples (numerical)
% Conclusion and Discussion

% Connection to known results for edge: only two kinds of gapped edge theories.
% ES allows us to understand gapped and gapless edge theories.

\section{Exact Wen-plaquette model}
\label{sect:Exact}
\subsection{Edge theory on semi-infinite cylinder}
\label{sect:edgeExact}
\begin{figure}\center
%\scalebox{.75}{\LatticeEdge}
\includegraphics[width=0.74\columnwidth]{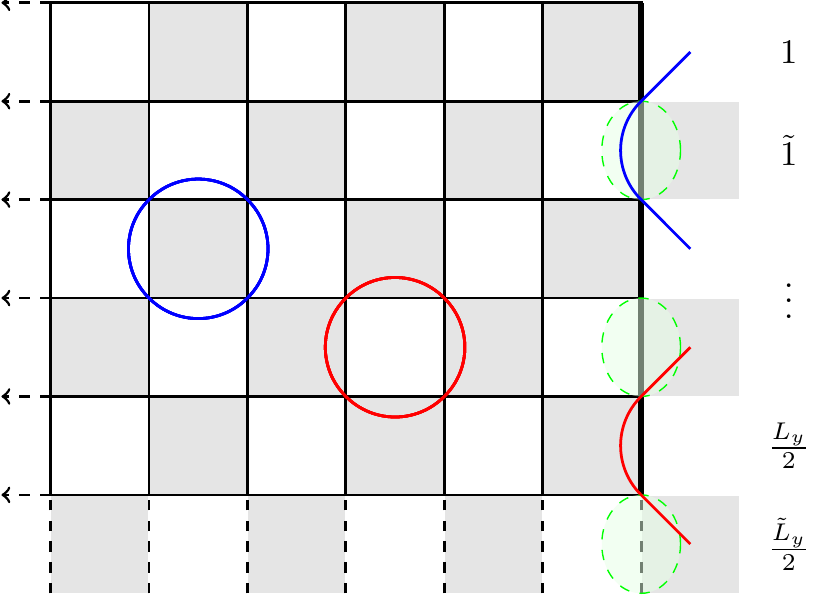}
\caption{(Color online.) Semi-infinite cylinder that terminates in the $x$-direction, with the periodic direction along $y$. The grey plaquette and white plaquette operators are shown as blue and red circles respectively. The boundary operators (BO) can be thought of as half of the plaquette operators in the bulk (acting on the white ($p$)  or grey (${\tilde p}$) plaquettes). The green ellipses on the edge correspond to the virtual, boundary spin degree of freedom. The blue BO acts on a single virtual spin, while the red BO acts on a pair of nearest-neighbor virtual spins.}
\label{fig:SemiInfiniteCylinder}
\end{figure}

We first consider the unperturbed $\Z_2$ Wen-plaquette model on a semi-infinite cylinder, $L$ (left), terminated with a smooth boundary, with the periodic direction along the $y$-axis. The Hamiltonian is the sum of commuting plaquette terms,
\begin{align}
H_L = - g \left( \sum_{\text{plaq.}} \mathcal{O}_\text{plaq.} - c \right) =  - \sum_\text{pl.} \Oop{\text{pl.}} - c,
\label{eqn:HL}
\end{align}
where $\mathcal{O}_\text{plaq.} = \scalebox{.8}{\OopNumb{\text{pl.}}} = Z_1 X_2 Z_3 X_4$ is the four-spin plaquette operator, and $\{X_i, Y_i, Z_i\}$ the set of Pauli matrices acting on site $i$. The energy scale $g$ has been set to $1$ and $c$ is a shift in energy such that the ground state energy of $H_L = 0$. The number of sites $L_y$ along $y$ is taken to be even, in order to avoid having a twist defect line (i.e. a consistent checkerboard coloring of the plaquettes can be made so that the elementary excitations $e$ and $m$ live on plaquettes of different colors). Figure \ref{fig:SemiInfiniteCylinder} shows the semi-infinite cylinder with a checkerboard coloring. 

Note that this is a choice of the Hamiltonian acting on the semi-infinite cylinder that we have made, as we need to also specify boundary conditions on the edge of the manifold. In particular, in \eqn{eqn:HL}, we have chosen free boundary conditions - we have simply taken the Wen-plaquette model on a semi-infinite cylinder to be the sum of plaquette operators with no additional commuting boundary Hamiltonian operators. The choice of different boundary conditions - the addition of boundary operators - will naturally affect the edge-ES correspondence, but we will only restrict our analysis to the case of free boundary conditions in this paper.

It will be useful to introduce the following graphical notation: we represent the Pauli operators as red string operators $Z_i =  \scalebox{.80}{\ZredLine{i}}$, $X_i = \scalebox{.80}{\XredLine{i}}$ if the strings live on the grey plaquettes $(\tilde{p})$; while we represent them as blue string operators $Z_i = \scalebox{.80}{\ZblueLine{i}}$, $X_i = \scalebox{.80}{\XblueLine{i}}$ if they live on the white plaquettes ($p$). For example, the {\it white} plaquette operator is given by $\scalebox{.80}{\OopRed{p}}$ since the strings live on the {\it grey} plaquettes neighboring it.

\vspace{5pt}
\noindent {\bf Topological sectors of ground state subspace.} The ground state subspace $\mathcal{V}_{0,L}$ of the Wen-plaquette model (defined by the plaquette condition $\scalebox{.75}{\Oop{\text{pl.}}} = + 1$) on an semi-infinite cylinder, similar to that of the infinite cylinder and of the torus, has four topological sectors $a$ ($\mathcal{V}_{0,L} = \oplus_{a = 1}^{N=4} \mathcal{V}_{0,L}^a$). They are distinguished by the eigenvalues of the two non-contractible Wilson loop operators around the cylinder $\tilde{\Gamma}_L = \scalebox{.55}{\GammaRed}_L$ and $ \Gamma_L=\scalebox{.55}{\GammaBlue}_L$, where $\tilde{\Gamma}_L$ is the string operator $Z_1 X_2 Z_3 \cdots X_{L_y}$ living on the grey plaquettes, acting on the spins on the boundary, as defined using the graphical notation above and similarly for $\Gamma_L$ for the white plaquettes. These operators commute with all $\mathcal{O}_{\text{plaq.}}$, square to $\mathds{1}$, and thus have eigenvalues $\tilde{Q}_L$ and $Q_L$ respectively taking values $\pm 1$ each, giving the four topological sectors $a \simeq (\tilde{Q}_L, Q_L)$. States within each topological sector $a$ are said to have well-defined anyonic flux with respect to $\tilde\Gamma_L$ and $\Gamma_L$.

\vspace{5pt}
\noindent {\bf Ground state degeneracy.}
The ground state subspace $\mathcal{V}_{0,L}$ is however not four-dimensional. To find the ground state degeneracy, we need to find a maximal set of commuting and independent operators in addition to the plaquette operators. Besides $\tilde\Gamma_L$ and $\Gamma_L$ we can have boundary operators (BO) $S_{p,L} = \scalebox{.75}{\EdgeRed{p}}_L; S_{\tilde p, L} = \scalebox{.75}{\EdgeBlue{\tilde p}}_L$, where $S_{p,L} = Z_{2p-1} X_{2p}$ and $S_{\tilde p, L} = Z_{2 \tilde{p}} X_{2 \tilde{p}+1}$ are half-plaquette operators or string operators acting on the spins on the boundary. The strings thus start and terminate outside the boundary of the semi-infinite cylinder (see Fig.~\ref{fig:SemiInfiniteCylinder}). There are $L_y/2$ of each kind of operators. The BOs individually commute with $\mathcal{O}_{\text{plaq.}}$, $\Gamma$ and $\tilde\Gamma$. However, $[S_{p,L}, S_{\tilde p,L}] \neq 0$ if $p$ neighbors $\tilde{p}$, so a choice of a maximal set of commuting and independent operators on the $L$ semi-infinite cylinder is 
\begin{align}
\left\{\mathcal O_p,\;\mathcal O_{\tilde p},\;S_{\tilde p},\;\tilde\Gamma\right\}_L  =  \left\{ \hspace{-5pt} \scalebox{.75}{ \OopRed{p}, \hspace{-2pt}\OopBlue{\tilde p},\;\EdgeBlue{\tilde p},\;\GammaRed} \right\}_L.
\end{align}
Note that $\Gamma_L$ is not included in the set because it can be formed by the boundary operators: $\Gamma_L = \prod_{\tilde p}  S_{\tilde p, L}$. Thus $\mathcal{V}_{0,L}$ has four topological sectors $a$, such that $\mathcal{V}_{0,L} = \oplus_{a = 1}^4 \mathcal{V}_{0,L}^a$, each with $L_y/2 - 1$ states labeled by the eigenvalues of $S_{\tilde p, L}$, for a total dimensionality $\text{dim}(\mathcal{V}_{0,L}) = 2^{L_y/2+1}$. Physically, these states can be thought of as having pairs of anyons ($e$ or $m$) that condense on the boundary - this process does not cost energy and thus the ground state degeneracy is given by the number of ways we can condense the anyons on the boundary. For this reason, these ground states can also be understood as edge states, and from now on, the terms `lowest-energy states' and `edge states' will be used interchangeably with `ground states'. We will also define the projector $P_{0,L}^a$ onto the eigenspace $\mathcal{V}_{0,L}^a$ for future use. 

At this point, we introduce a mapping of the boundary operators $S_{p,L}$ and $S_{\tilde{p},L}$ to local operators acting on a finite length (spin-$1/2$) spin chain. This mapping will be important as it elucidates the tensor product structure of the edge theory on the semi-infinite cylinder.  Consider the Wilson loop operator $\scalebox{.55}{\GammaRed}_L$ partitioning $\mathcal{V}_{0,L}$ into two: $\mathcal{V}_{0,L} = \oplus_{\tilde{Q} = \pm 1} \mathcal{V}_{0,L}^{\tilde{Q}_L}$, labeled by $\tilde{Q}_L$.  Each $\tilde{Q}_L$ sector has dimension $2^{L_y/2}$, which is isomorphic to a spin-$1/2$ chain of length $L_y/2$.  Let us therefore associate a virtual spin-$1/2$ degree of freedom for each $\tilde{p}$-plaquette (see Fig.~\ref{fig:SemiInfiniteCylinder}). Here, we see that a (not unique) representation of the operators $S_{p,L}$ and $S_{\tilde p,L}$ acting on the $L$ spin-$1/2$ chain for each $\tilde{Q}_L$ can be found:
\begin{align}
&\scalebox{.75}{\EdgeBlue{\tilde p}} \simeq \tau_{\tilde{p},L}^x \text{ for } 1 \leq \tilde{p} \leq \tilde{L}_y/2, \nonumber \\
&\scalebox{.75}{\EdgeRed{p}} \simeq \tau_{\tilde{p},L}^z \tau_{\tilde{p}-1,L}^z \text{ for } 2 \leq \tilde{p} \leq \tilde{L}_y/2,
\label{eqn:Map}
\end{align}
and $\scalebox{.75}{\EdgeRed{1}}  \simeq \tilde{Q}_L \times \tau_{\tilde{1},L}^z \tau_{\tilde{L}_y/2,L}^z$, i.e. toroidal boundary conditions. One can check that the Pauli spin operators reproduce the canonical anticommutation algebra of $S_{p,L}$ and $S_{\tilde p,L}$ and that $\scalebox{.55}{\GammaRed}_L = \tilde{Q}_L$ is satisfied. Note that the Wilson loop $\scalebox{.55}{\GammaBlue}_L$ is mapped to the global spin-flip operator $\hat{Q}_L := \prod_{\tilde p} \tau_{{\tilde p},L}^x$, with eigenvalues $Q_L = \pm 1$. A similar representation can be found for the operators $S_{p,R}$ and $S_{\tilde{p},R}$ acting on the spin-$1/2$ chain of $R$.

\vspace{5pt}
\noindent {\bf Higher energy subspaces.}
Higher energy subspaces $\mathcal{V}_{\alpha > 0,L}$ are spanned by states for which the plaquette condition $\scalebox{.75}{\Oop{\text{pl.}}} = + 1$ is violated. As such, there is a spectral gap of at least $+1$ separating the ground state subspace from the higher energy subspaces. These violations are generated by string operators that have at least one end point in the bulk.  

For each higher subspace $\mathcal{V}_{\alpha > 0,L}$, we can further define a notion of topological sectors $a$, where $a = 1,\cdots 4$, in the following way: the subspace $\mathcal{V}_{\alpha >0, L}^{a}$, is the space spanned by all states in $\mathcal{V}_{0,L}^a$ which are acted upon by all possible products of finite-length (i.e. local) string operators such that the number of end-points of these string operators in the bulk of $L$ is $\alpha$.   These sectors are called topological because the matrix element of generic local operators using states belonging to different topological sectors vanishes.

\vspace{5pt}
\noindent{\bf Edge theory.}
The edge theory in each topological sector $a$ is defined to be the Hamiltonian $H_{\text{edge},L}^a$ acting on the $2^{L_y/2- 1}$ edge states of the subspace $\mathcal{V}_{0,L}^a$ that give rise to the different states' energy levels. However, all states in $\mathcal{V}_{0,L}$ have the exact same energy, and hence the edge Hamiltonian for the exact Wen-plaquette model in each topological sector is identically $0$. 

\vspace{5pt}
Let us summarize what we have learned:
\vspace{5pt}

\noindent
$\bullet$ 
The effective low energy degrees of freedom ($\mathcal{V}_{0,L}^{\tilde{Q}_L}$) at the boundary of the $L$  semi-infinite cylinder of length $L_y$ is a spin chain made of $L_y/2$ virtual spin-$1/2$ degrees of freedom, for each topological $\tilde{Q}_L$ sector. The spin chain is generated by the boundary operators $S_{p,L}$ and $S_{\tilde p,L}$ which are half-plaquette operators in the bulk. In the effective spin chain language, these boundary operators are mapped to $\mathbb{Z}_2$ symmetric spin-operators $\tau_{\tilde{p},L}^z \tau_{\tilde{p}-1,L}^z$ and $\tau_{\tilde{p},L}^x$. Without perturbations, $H_{\text{edge},L}^a$ is identically $0$. With perturbations, there will be dynamics on this effective spin chain, generated by these boundary operators. A similar situation arises for the $R$ semi-infinite cylinder.

\subsection{Entanglement spectrum}
\label{sect:unpertES}
\begin{figure}\center
\includegraphics[width=1\columnwidth]{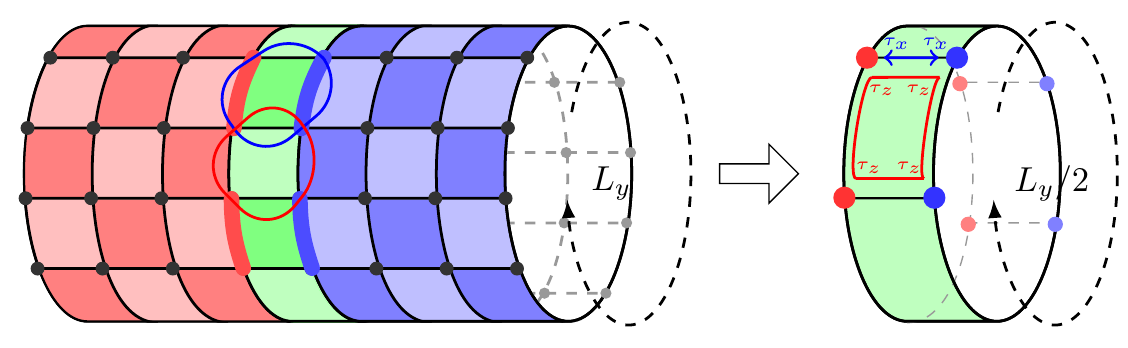}
\caption{(Color online). (Left) Gluing two semi-infinite cylinders of circumference $L_y$ together. The $L$ semi-infinite cylinder is colored red, the $R$ semi-infinite cylinder is colored blue, while the plaquette terms belonging to the strip are colored green. The entanglement cut is naturally taken to be the divisor through the strip so that the system is bipartitioned into the $L$ and $R$ semi-infinite cylinders. A checkerboard coloring has been made on the infinite cylinder, and the $L_y/2$ effective virtual spin-$1/2$ degrees of freedom are denoted by red and blue ellipses living on the boundaries of the semi-infinite cylinders.  (Right) The mapping, \eqn{eqn:Map}, gives rise to an effective Hamiltonian acting on the spin-ladder system, with each spin chain having $L_y/2$ sites, which were previously the red and blue ellipses. The red and blue plaquette operators have been mapped to $z$-rung and $x$-rung operators respectively. The effective Hamiltonian's ground states correspond to the ground states of the infinite cylinder.}
\label{fig:cylinder}
\end{figure}

Let us now solve for the four ground states of the exact Wen-plaquette model on the infinite cylinder, and compute their entanglement spectrum for a bipartition of the infinite cylinder into two semi-infinite cylinders. We can do this by putting two semi-infinite cylinders $L$ and $R$ together, and gluing them with the plaquette terms that act on the strip of plaquettes spanning the two cylinders. That is, we solve:
\begin{align}
H = H_L + H_R + H_{LR},
\label{eqn:unpertH}
\end{align}
where $H_L$ is given by \eqn{eqn:HL} (and correspondingly for $H_R$), acting on the $L$ and $R$ semi-infinite cylinders respectively, while $H_{LR} = -\sum_{\text{plaq.} \in \text{strip}} \mathcal{O}_{\text{plaq.}}$. The entanglement cut is naturally taken to be through the strip of plaquettes that divides the system into the $L$ and $R$ subsystems, such that the full Hilbert space $\mathcal{H}$ is the tensor product of the two semi-infinite cylinders: $\mathcal{H} = \mathcal{H}_L \otimes \mathcal{H}_R$. Figure \ref{fig:cylinder} shows the gluing process.

Now, $\mathcal{H}$ can also be written as $\mathcal{H} = \bigoplus_{\alpha,\beta = 0}^\infty \mathcal{V}_{\alpha,L} \otimes \mathcal{V}_{\beta,R}$. Note that a plaquette operator in $H_{LR}$ is comprised of two matching boundary operators acting on the $L$ and $R$ cylinders, for example $\mathcal{O}_{\text{plaq.}\in\text{strip.}}(p) = S_{p,L} \otimes S_{p,R}$ (note that $p$ is the same for $L$ and $R$) and similarly for $\tilde{p}$, which implies that its action must be such that the tensor product structure is preserved, $H_{LR} :\mathcal{V}_{\alpha,L} \otimes \mathcal{V}_{\beta,R} \to \mathcal{V}_{\alpha,L} \otimes \mathcal{V}_{\beta,R}$. Since $H_{LR}$ is the sum of mutually commuting terms, we can simply focus on its action on the $ \mathcal{V}_{0,L} \otimes \mathcal{V}_{0,R}$ sector because that is where the ground states of the infinite cylinder live in.

Next, we employ the fact that there are topological sectors $a$ and $a'$ in $\mathcal{V}_{0,L}$ and $\mathcal{V}_{0,R}$ to further narrow down the subspace of the Hilbert space where the ground states reside in. Consider the subspace given by tensoring subspaces of $L$ and $R$ with $\tilde{Q}_L \neq \tilde{Q}_R$: $\mathcal{V}_{0,L}^{\tilde{Q}_L} \otimes \mathcal{V}_{0,R}^{\tilde{Q}_R}$ . In this space, $\scalebox{.55}{\GammaRed}_L \otimes \scalebox{.55}{\GammaRed}_R = -1$, which means that one of the plaquette conditions of the plaquettes in $H_{LR}$ is violated. Thus, the ground states cannot live in this space. Conversely it means that the ground states must have $\tilde{Q}_L = \tilde{Q}_R \equiv \tilde{Q}$.

A similar argument can be made for $\scalebox{.55}{\GammaBlue}_L$ (with eigenvalues $Q_L = \pm1$), which would yield $Q_L = Q_R \equiv Q$ for the ground states. This would imply that $a = a'$ for the ground states (or equivalently $a \simeq (\tilde{Q}_L, Q_L) = (\tilde{Q}_R,Q_R) \simeq a'$ as there is a one-to-one map between $a$ and $(\tilde{Q}_L,Q_L)$). Let us therefore first consider projecting $H_{LR}$ into the $\tilde{Q}$-sector subspace $\mathcal{V}_{0,L}^{\tilde{Q}} \otimes \mathcal{V}_{0,R}^{\tilde{Q}}$ by the projector $P_{0,L}^{\tilde{Q}} \otimes P_{0,R}^{\tilde{Q}}$. 

Using the mapping, \eqn{eqn:Map}, one possible representation of the strip-plaquette operators upon projection into $\mathcal{V}_{0,L}^{\tilde{Q}} \otimes \mathcal{V}_{0,R}^{\tilde{Q}}$, is of operators acting on two spin chains (of length $L_y/2$ each) for the left and right cylinders:
\begin{align}
&\mathcal{O}_{\tilde{p}} = \scalebox{.75}{\OopBlue{\tilde{p}}} \to \tau_{ \tilde{p},L}^{x} \tau_{\tilde{p},R}^{x} \nonumber\\
& \mathcal{O}_{p} = \scalebox{.75}{\OopRed{p}} \to \tau_{\tilde{p},L}^{z} \tau_{\tilde{p}-1,L}^{z} \tau_{\tilde{p},R}^{z} \tau_{\tilde{p}-1,R}^{z},
\end{align}
for both $\tilde{Q}$. These operators are called ``$x$-rung'' and ``$z$-rung'' operators respectively.

The effective Hamiltonian, i.e. the projection of $H_{LR}$ into $\mathcal{V}_{0,L}^{\tilde{Q}} \otimes \mathcal{V}_{0,R}^{\tilde{Q}}$, is therefore 
\begin{align}
H_{\text{eff}} = - \sum_{\tilde{p} = 1}^{\tilde L_y/2} (\tau_{\tilde{p},L}^{x} \tau_{\tilde{p},R}^{x} +  \tau_{\tilde{p},L}^{z} \tau_{ \tilde{p}-1,L}^{z} \tau_{\tilde{p},R}^{z} \tau_{\tilde{p}-1,R}^{z}).
\label{eqn:EffEntH}
\end{align}
This is a spin-ladder system (see Fig.~\ref{fig:cylinder}), of length $L_y/2$, with $\mathcal{O}_p$ and $\mathcal{O}_{\tilde{p}}$ coupling the two chains. It also has a $\mathbb{Z}_2$ symmetry generated by the global spin flip operator $\hat{Q}_L := \prod_{\tilde{p}} \tau_{\tilde{p},L}^x$ (whose eigenvalue is $Q_L$). Solving \eqn{eqn:EffEntH} for both $\tilde{Q}$ will give us all four ground states of the system.

Let us fix a $\tilde{Q}$. Define the (unique) state $\ket{\uparrow}_L$ as the state that is a ground state of $H_L$ with $\tilde{Q}_L = \tilde{Q}$ and furthermore satisfies $\tau_{\tilde{p}, L}^z = +1$. Then we see that the other $2^{\tilde{L}_y/2}$ states spanning $\mathcal{V}_{0,L}^{\tilde{Q}}$ comprise of all other possible spin configurations $\ket{\tau}_L$, generated by $\tau_{\tilde{p},L}^x$ acting on $\ket{\uparrow}_L$. The same is also true for $R$. The ground state of \eqn{eqn:EffEntH}, or equivalently of the full Hamiltonian, in this $\tilde{Q}$ sector with $Q = Q_L = Q_R$ is therefore given by
\begin{align}
\ket{\tilde{Q},  Q}\hspace{-3pt}  &= \hspace{-3pt} \mathcal{N} \prod_{\tilde{p}} \left( \frac{\mathds{1} + \tau_{\tilde{p},L}^x \tau_{\tilde{p},R}^x }{2} \right) \left( \ket{\uparrow}_L \ket{\uparrow}_R + Q \hat{Q}_L \ket{\uparrow}_L \ket{\uparrow}_R \right) \nonumber \\
& = \frac{1}{\sqrt{2^{\tilde L_y/2}}} \frac{1}{\sqrt{2}} \sum_{\tau} \left( \ket{\tau}_L \ket{\tau}_R + Q \ket{\bar{\tau}}_L \ket{\tau}_R \right) \nonumber \\
& =  \frac{1}{\sqrt{2^{\tilde L_y/2 - 1}}} \sum_{\tau}  \mathbb{P}_{Q,L} \ket{\tau}_L \ket{\tau}_R,
\label{eqn:ZOGS}
\end{align}
where $\mathbb{P}_{Q,L} := (\mathds{1} + Q_L \hat{Q}_L)/2$ is the $\mathbb{Z}_2$ projector onto the $Q_L = Q$ sector in the $L$ chain. Thus, in total, there are four ground states on the infinite cylinder as claimed, each with well-defined anyonic flux through the cylinder.

Lastly, from \eqn{eqn:ZOGS}, it can be readily seen that the entanglement spectrum of $\ket{\tilde{Q},Q}$ is flat: the reduced density matrix on $L$ is
\begin{align}
\rho_L^{a\simeq (\tilde{Q},Q) } := \Tr_R \ket{\tilde{Q},Q}\bra{\tilde{Q},Q} = \frac{1}{2^{\tilde{L}_y/2 - 1}} \mathbb{P}_{Q,L},
\label{eqn:RDM}
\end{align}
with $2^{\tilde{L}_y/2 - 1}$ non-zero eigenvalues all of value $1/2^{\tilde{L}_y/2 - 1}$. The entanglement Hamiltonian $H_{\text ent.}^a \equiv -\ln(\rho_L^a)$ is therefore also flat and acts on a virtual spin chain of length $L_y/2$.

\vspace{5pt}
In summary: 
\vspace{5pt}

\noindent 
$\bullet$ 
The entanglement Hamiltonian of the exact Wen-plaquette model in topological sector $a$ is flat and acts on a virtual spin-$1/2$ spin chain of length $L_y/2$, similar to the case of the edge Hamiltonian. Each entanglement Hamiltonian gives an ES with $2^{L_y/2-1}$ finite values. The entanglement Hamiltonians can be derived by considering an effective Hamiltonian acting on a spin-ladder system with length $L_y/2$, for each $\tilde{Q} = \tilde{Q}_L = \tilde{Q}_R$ sector. Solving the effective Hamiltonian in each $\tilde{Q}$ sector gives two ground states distinguished by the eigenvalue $Q = Q_L = Q_R$, for a total of four ground states overall.

\subsection{Edge-ES correspondece}
Since the edge Hamiltonian is identically $0$, and the entanglement Hamiltonian flat, the edge-ES correspondence is exact in this case: the two spectra are equal up to rescaling and shifting. However, that is not to say that the flat entanglement spectrum is uninteresting: for example, the entanglement entropy in each topological sector can be readily calculated, yielding
\begin{align}
S_a = \Tr(\rho_L^a \ln \rho_L^a ) = \left( \tilde L_y /2 - 1 \right) \ln 2.
\end{align}
From there the topological entanglement entropy\cite{LevinWenEE2006,KitaevPreskillEE2006} $\gamma_a$, defined to be the universal sub-leading piece of the entanglement entropy, $S_a = \alpha_a L - \gamma_a + \cdots$, can be extracted:
\begin{align}
\gamma_a = \ln 2,
\end{align}
in agreement with the fact that $\ket{\tilde{Q},Q}$ are the so-called minimum entangled states\cite{ZhangGroverOshikawaVishwanath}  on the infinite cylinder.
%Vishwanath MES

\vspace{5pt}
We conclude:
\vspace{5pt}

\noindent
$\bullet$ 
The edge-ES correspondence for the exact Wen-plaquette model is exact: all $2^{L_y/2 - 1}$ levels of the edge and entanglement spectra in a topological sector $a$ conincide with a shift and rescaling that is common to all sectors $a$.

%%%%%%%%%%%%%%%%%%%%

\section{Perturbed Wen-plaquette model}
\label{sect:Perturbed}

In this section, we perturb the Wen-plaquette model and derive both the edge theory on the semi-infinite cylinder and entanglement spectrum of the ground states on the infinite cylinder. We shall be precise as to what we mean by the edge theory in this case. The general perturbed Wen-plaquette model on the full cylinder is
\begin{align}
H = H_L + H_R + H_{LR} + \epsilon(V_L + V_R + V_{LR}),
\label{eqn:pertH}
\end{align}
where $H_L + H_R + H_{LR}$ is as before, in \eqn{eqn:unpertH}, while $V_L$ and $V_R$ are perturbations acting on each respective semi-infinite cylinder, and $V_{LR}$ is a perturbation that spans the cut. We assume that for weak enough perturbations, the lowest energy subspaces of $H_L$ and $H_L + \epsilon V_L$ are adiabatically connected. Thus, energy levels in the highly degenerate subspace $\mathcal{V}_{0,L}$ of $\mathcal{H}_L$ acquire dispersions due to the perturbations, and split. The edge theory or edge Hamiltonian $H_{\text{edge},L}$ is then defined to be the Hamiltonian acting on the states in the lowest energy eigenspace of $H_L + \epsilon V_L$ that generates the dynamics and hence the dispersion.

The key to deriving this edge Hamiltonian and subsequently, the entanglement spectrum of the four ground states of \eqn{eqn:pertH}, is the Schrieffer-Wolff (SW)\cite{SchriefferWolff} transformation. This is a transformation which perturbatively block diagonalizes a Hamiltonian if its original, unperturbed Hamiltonian was block-diagonal to begin with, which is the case for the Wen-plaquette model. We begin this section by introducing the SW transformation before then applying it to finding the edge and the entanglement Hamiltonians.

\subsection{Mathematical preliminaries: Schrieffer-Wolff transformation}
\label{sect:SW}
We present a concise but necessary introduction to the Schrieffer-Wolff (SW) transformation\cite{SchriefferWolff}. See \Ref{BravyiDiVincenzoLoss} for all precise definitions and theorems concerning the SW transformation. 
% Bravyi, DiVincenzo & Loss

Let us be given a Hamiltonian $H_0$ that has a low-energy eigenspace $\mathcal{V}_0$ and a high-energy eigenspace $\mathcal{V}_1$ separated by a spectral gap. Then $H_0$ can be written as
\begin{align}
H_0 = P_0 H_0 P_0 + P_1 H_0 P_1 ,
\end{align}
where $P_\alpha$ are projectors to $\mathcal{V}_\alpha$, $\alpha = 0,1$. Let us now add a small perturbation to the system $\epsilon V$ that does not commute with $H_0$. We assume that the perturbations are weak enough such that the new Hamiltonian can be written as
\begin{align}
H = H_0 + \epsilon V = \tilde{P}_0 H \tilde{P}_0 + \tilde{P}_1 H \tilde{P}_1,
\end{align}
where there are still low-energy $\mathcal{\tilde{V}}_0$ and high-energy $\mathcal{\tilde{V}}_1$ subspaces separated by the spectral gap. $\mathcal{V}_\alpha$ and $\mathcal{\tilde{V}}_\alpha$ are assumed to have the same dimensionality, and have significant overlap. Then, we can find a unique direct rotation (i.e. unitary) $U$ between the old and new subspaces ($ U \tilde{P}_\alpha U^\dagger = P_\alpha$) such that we can rotate $H$ to a new Hamiltonian $H'$ with eigenspaces $V_\alpha$:
\begin{align}
H' := U H U^\dagger = P_0 U H U^\dagger P_0  + P_1 U H U^\dagger P_1.
\end{align}
This is the so-called Schrieffer-Wolff transformation\cite{SchriefferWolff, BravyiDiVincenzoLoss}. It will be useful that $U$ can be written uniquely as $U = e^S$, where $S$ is an antihermitian and block off-diagonal (in both $\mathcal{V}_\alpha$ and $\mathcal{\tilde{V}}_\alpha$) operator, and can be constructed perturbatively in $\epsilon$: $S = \sum_{n \geq 1} \epsilon^n S_n$, $S_n^\dagger = -S_n$. The exact formulas for $S_n$ can be found in \Ref{BravyiDiVincenzoLoss}, and we reproduce them in Appendix \ref{sect:AppendixA}.

Though our discussion above has been limited to Hamiltonians with only two invariant subspaces (low and high), the SW transformation can be readily generalized to Hamiltonians that have many invariant subspaces each separated by a spectral gap, such that the Hilbert space $\mathcal{H} = \bigoplus_{\alpha \geq 0} \mathcal{V}_\alpha$, see \Ref{Zheng}. $S$ will still be block-off-diagonal.
%http://journals.aps.org/prb/pdf/10.1103/PhysRevB.63.144410

The generalized Schrieffer-Wolff transformation is also referred to as the effective Hamiltonian method\cite{SachdevBook}. To second order in perturbation theory, $H'$, which by construction is block-diagonal in $\mathcal{V}_\alpha$, is given explicitly by
\begin{align}
&\bra{i,\alpha}H'\ket{j,\alpha} = E^\alpha_i \delta_{ij} + \epsilon \bra{i,\alpha}V\ket{j,\alpha}   + \frac{\epsilon^2}{2} \sum_{\substack{k \\ \beta \neq \alpha} } \nonumber \\
& \bra{i,\alpha}V\ket{k,\beta} \bra{k, \beta}V\ket{j,\alpha}\left( \frac{1}{E_i^\alpha - E_k^\beta} + \frac{1}{E_j^\alpha - E_k^\beta} \right),
\label{eqn:effH}
\end{align}
where $\ket{i,\alpha} \in \mathcal{V}_\alpha$, $\ket{j,\beta} \in \mathcal{V}_\beta$ and $E_{i}^\alpha$ is the energy of $\ket{i,\alpha}$. See appendix \ref{sect:AppendixA} for the full perturbative series of the effective Hamiltonian.
%Sachdev

\subsection{Edge theory on semi-infinite cylinder}
\label{sect:pertEdge}
As mentioned before, the SW transformation is suitable for use on the perturbed Wen-plaquette model on the semi-infinite cylinder,  $H_L + \epsilon V_L$, because the unperturbed Hamiltonian $H_L$ is block-diagonal with spectral gaps separating the different energy eigenspaces. To find the edge theory, $H_{\text{edge},L}$, we want to evaluate \eqn{eqn:effH} (with $V \to V_L$) for the $L$ semi-infinite cylinder with states belonging to $\mathcal{V}_{0,L}$, the lowest energy subspace. Furthermore, we can make use of the fact that states with different $\tilde{Q}_L$ do not mix at any order in perturbation theory, as the perturbations are local and cannot generate a global term that mixes $\tilde{Q}_L$ sectors, so we can consider \eqn{eqn:effH} restricted to states belonging to $\mathcal{V}_{0,L}^{\tilde{Q}_L}$.

However, since by construction $H_{\text{edge},L}^{\tilde{Q}_L}$ acts only on $\mathcal{V}_{0,L}^{\tilde{Q}_L}$, $H_{\text{edge},L}^{\tilde{Q}_L}$ is generated solely from virtual processes in the perturbative series (\eqn{eqn:effH}) that map states in the ground state subspace back to itself. These virtual processes are simply products of boundary operators and plaquette operators on $L$.

To clarify this statement, let us show this for a particular second order term in a fixed $\tilde{Q}_L$ sector. Let $v_{1,L}$ and $v_{2,L}$ be two local perturbations, each of which are not sums of perturbations, coming from $V_L$ (which is generally a sum of local perturbations) that give rise to a non-zero contribution in the matrix element, i.e., $\bra{i,0}v_{1,L}\ket{k,\beta} \bra{k, \beta}v_{2,L}\ket{j,0} \neq 0$ for some $i,j,k,\beta$. $v_{2,L}$ is comprised of products of string operators with end points in the bulk of the $L$ semi-infinite cylinder. The number of end points in $L$ determines $\beta$, i.e. $v_{2,L}$ can link a ground state to a state with $\beta$ excitations in $L$. This is a state which lives in $\mathcal{V}_{\beta,L}$. Furthermore, this state is unique. Indeed, $\ket{k,\beta} = v_{2,L}\ket{j,0}$, and so $\bra{l, \gamma}v_{2,L}\ket{j,0} = 0 $ for any other state such that $\ket{l,\gamma} \neq \ket{k,\beta}$. Now, since $\bra{i,0}v_{1,L} \ket{k,\beta} \neq 0$, $v_{1,L}$ must be such that it creates the same bulk excitations as $v_{2,L}$, in order to cancel out the excitations of $\ket{k,\beta}$, modulo products of plaquette and boundary operators. The presence of the boundary operators makes $\ket{i,0}$ potentially not equal to $\ket{j,0}$. Putting all these facts together, we see that replacing $\ket{k,\beta}\bra{k,\beta}$ with the sum over all possible states in the Hilbert space, i.e. $\sum_{l,\gamma} \ket{l, \gamma}\bra{l, \gamma} = \mathds{1}$ for this $v_{1,L}, v_{2,L}$ does not change the value of the matrix element, yielding the desired assertion:
\begin{align}
&\bra{i,0}v_{1,L}\ket{k,\beta} \bra{k, \beta}v_{2,L}\ket{j,0} = \bra{i,0}v_{1,L} \mathds{1} v_{2,L}\ket{j,0} = \nonumber \\
& \bra{i,0}
\hspace{-10pt} \prod_{ \substack{ \{\text{pl.} ,p,\tilde{p} \} \\ \text{ of } v_{1,L}v_{2,L} } } \hspace{-10pt} \scalebox{.75}{\Oop{\text{pl.}}} \times \scalebox{.75}{\EdgeRed{p} } \hspace{-5pt} \times   \scalebox{.75}{\EdgeBlue{\tilde p} } 
\ket{j,0}.
\end{align}
A similar argument can be made for terms of other orders in perturbation theory.

%%%%%%%%%%%%%%%%
\begin{comment}
We claim that the only dynamics of $H_{\text{edge},L}^{\tilde{Q}_L}$ comes about when the products of $V_L$ in \eqn{eqn:effH} are products of boundary operators $S_{p}$, $S_{\tilde{p}}$ modulo products of plaquette operators $\mathcal{O}_\text{plaq.}$. This is because these are the only operators that can connect two states in $\mathcal{V}_{0,L}^{\tilde{Q}_L}$. For example, in second order in perturbation theory, the only possible non-zero matrix element is if (schematically)
\begin{align}
 &\bra{i,0}V_L\ket{k,\beta} \bra{k, \beta}V_L\ket{j,0} \nonumber \\ 
=  &\sum_{\mu} c_\mu \bra{i,0} \hspace{-10pt} \prod_{ \{\text{pl.} ,p,\tilde{p} \} \in \mu} \hspace{-10pt} \scalebox{.75}{\Oop{\text{pl.}}} \times \scalebox{.75}{\EdgeRed{p} } \hspace{-5pt} \times   \scalebox{.75}{\EdgeBlue{\tilde p} }  \ket{j,0},
\label{eqn:prodBO}
\end{align}
and the same is true for other orders in perturbation theory. Here, the sum over $\mu$ corresponds to the sum over all non-zero contributons arising from $V_L$ (which is itself a sum of perturbations). $\mu$ also labels a set containing a sequence of plaquettes (pl.), a sequence of $p$, and a sequence of $\tilde{p}$ which a non-zero contribution maps to.
\end{comment}
%%%%%%%%%%%%%%%%%%%%

Since $\scalebox{.75}{\Oop{\text{pl.}}} = +1$ for the ground states, we therefore see that $H_{\text{edge},L}^{\tilde{Q}_L}$ must be a function of products of $S_{p,L}$ and $S_{\tilde{p},L}$ projected down into $\mathcal{V}_{0,L}^{\tilde{Q}_L}$:
\begin{align}
H_{\text{edge},L}^{\tilde{Q}_L} = f_{\tilde{Q}_L,L} \bigg(\scalebox{.75}{\EdgeRed{p} }_L \hspace{-5pt},  \scalebox{.75}{\EdgeBlue{\tilde p} }_L \bigg).
\end{align}
However, recall the mapping given by \eqn{eqn:Map}, which allows us to interpret the edge Hamiltonian in terms of a more physical picture: a Hamiltonian acting on a spin chain.

The edge Hamiltonian in a $\tilde{Q}_L$-sector at finite order in perturbation theory is therefore a local Hamiltonian acting on a spin-$1/2$ chain of length $L_y/2$:
\begin{align}
H_{\text{edge},L}^{\tilde{Q}_L} = f_{\tilde{Q}_L,L}(\tau_{\tilde{p},L}^x, \tau_{\tilde{p},L}^z \tau_{\tilde{p}-1,L}^z),
\label{eqn:edgeHtildeQ}
\end{align}
with toroidal boundary conditions given by $\tilde{Q}_L$ (i.e. $\tau_{\tilde{0},L}^z = \tilde{Q} _L\tau_{\tilde{L}_y/2,L}^z$). Since it is constructed perturbatively, it generically appears at order $\epsilon$. Importantly, we also see that the edge Hamiltonian is always $\mathbb{Z}_2$ symmetric regardless of the type of perturbation. This is not a surprising result because the two sectors of $\mathbb{Z}_2$ in the spin language, given by the generator $\hat{Q}_L := \prod_{\tilde p} \tau_{{\tilde p},L}^x$, correspond to the two topological sectors $Q_L = \pm 1$ of the Wilson loop $\scalebox{.55}{\GammaBlue}_L$, which are preserved under any local perturbations. The edge Hamiltonian in each topological sector $a$ then arises from projecting $H_{\text{edge},L}^{\tilde{Q}_L}$ into the relevant $Q$ sector: $H_{\text{edge},L}^a \equiv \mathbb{P}_{Q,L} H_{\text{edge},L}^{\tilde{Q}_L} \mathbb{P}_{Q,L}$.

\vspace{5pt}
In summary:
\vspace{5pt}

\noindent 
$\bullet$ 
The edge Hamiltonian of the perturbed Wen-plaquette model for the $L$ semi-infinite cylinder is given by $H_{\text{edge},L}^{\tilde{Q}_L}$, \eqn{eqn:edgeHtildeQ}, in each $\tilde{Q}_L $ sector. At any finite order in perturbation theory, it is a $\mathbb{Z}_2$ symmetric local Hamiltonian on the virtual spin chain. This is guaranteed if the perturbations of the Wen-plaquette model are themselves local on the cylinder. To further find the edge Hamiltonians in each topological sector $a \simeq (\tilde{Q}_L,Q_L)$, we project into the $Q_L$ sector:
\begin{align}
H_{\text{edge},L}^a \equiv \mathbb{P}_{Q,L} H_{\text{edge},L}^{\tilde{Q}_L}  \mathbb{P}_{Q,L},
\label{eqn:edgeH}
\end{align}
where $\mathbb{P}_{Q,L} = (\mathds{1} + Q_L \hat{Q}_L )/2$.  A similar result holds for the $R$ semi-infinite cylinder.

\subsection{Entanglement spectrum}
\label{sect:pertES}
We now find the ground states in each topological sector $a$ of the perturbed Wen-plaquette model on the infinite cylinder,
\begin{align}
H = H_L + H_R + H_{LR} + \epsilon(V_L +  V_R + V_{LR}).
\end{align}
The precise definition of each term can be found in \eqn{eqn:pertH}.

Let us fix a sector $a \simeq (\tilde{Q},Q)$. There will now be corrections to the ground state leading to a ES with dispersion. However, there now exists a difficulty in comparing the edge and entanglement spectra. From \eqn{eqn:edgeH}, we see that $H_{\text{edge}}^a$ still has $2^{L_y/2-1}$ eigenvalues; but, there will be many more entanglement energies than the $2^{L_y/2-1}$ finite ones as found in \eqn{eqn:ZOGS}. What does it mean to compare the two spectra then? The resolution can be found by looking at the general form of the perturbed ground state. Dropping the normalization constant for now, for a fixed $\tilde{Q}$, the perturbed ground state can be written as
\begin{align}
\ket{\tilde{Q}, Q} = & \sum_{\tau',\tau} ( \mathbb{P}_Q -\Lambda)_{\tau',\tau} \ket{\tau'}_L \ket{\tau}_R + \nonumber \\
& \sum_{\tau, i, \alpha \geq 1} \left( \Theta_{\tau, \{i,\alpha \} } \ket{\tau}_L \ket{i,\alpha}_R + \Xi_{\{i,\alpha \},\tau } \ket{i, \alpha}_L \ket{\tau}_R \right) \nonumber \\
& + \sum_{\substack{i,\alpha\geq 1\\j,\beta\geq1}} \Omega_{\{i,\alpha\},\{j,\beta\}} \ket{i,\alpha}_L \ket{j,\beta}_R,
\label{eqn:genericState}
\end{align}
where $(\mathbb{P}_Q)_{\tau', \tau}$ are the matrix elements of the matrix $\mathbb{P}_Q =  (\mathds{1} + Q \prod_i \tau^x_i )/2$, written in the $\tau^z$ basis. $\Lambda, \Theta, \Xi, \Omega$ are coefficient matrices linking a state in $L$ with another state in $R$ and are small in magnitude ($\epsilon$ or higher). Here $\ket{\tau'}_L \in \mathcal{V}_{0,L}^{\tilde{Q}}$, and $\ket{i,\alpha} \in \mathcal{V}_{\alpha,L}$ for $\alpha \geq 1$. The same holds true for states in $R$.

We have written the ground state in a way to emphasize the change in the entanglement structure. Our claim is that for generic local perturbations, $\Lambda$ has the form
\begin{align}
\Lambda = \mathbb{P}_Q \Lambda' \mathbb{P}_Q,
\end{align}
which is an order $\epsilon$ correction which is related to $H_{\text{ent.}}^a$. Here we have implicitly defined the central object of interest, $\Lambda'$, which we will argue is related to the entanglement Hamiltonian.

Furthermore, $\Lambda$ is generated by the terms $H_{\text{edge},L}^a$, $H_{\text{edge},R}^a$, and $(P_{0,L}^a \otimes P_{0,R}^a ) V_{LR} (P_{0,L}^a \otimes P_{0,R}^a)$. We see that $\mathbb{P}_Q \Lambda' \mathbb{P}_Q$ is a $2^{L_y/2}$ by $2^{L_y/2}$ matrix with $2^{L_y/2-1}$ eigenvalues that have eigenvectors with $\mathbb{P}_Q = 1$, linking states in $ \mathcal{V}_{0,L}^{\tilde{Q}}$ with $\mathcal{V}_{0,R}^{\tilde{Q}}$ which is spanned by $\ket{\tau'}_L \otimes \ket{\tau}_R$. $\Theta$ and $\Xi$ are generically order $\epsilon$ corrections and links states in $\mathcal{V}_{0,L}^{\tilde{Q}}$ with $\mathcal{V}_{\alpha\geq0,R}$ and $ \mathcal{V}_{\alpha\geq0,L}$ with $\mathcal{V}_{0,R}^{\tilde{Q}}$ respectively. $\Omega$ represents order $\epsilon$ corrections to the ground state in the space $\mathcal{V}_{\alpha\geq0,L} \otimes\mathcal{V}_{\beta\geq0,R}$. 

We sketch here why $\Lambda = \mathbb{P}_Q \Lambda' \mathbb{P}_Q$ is related to the entanglement Hamiltonian. If we form the un-normalized reduced density matrix on $L$, $\rho_L^a = \Tr_R \ket{\tilde{Q},Q}\bra{\tilde{Q},Q}$, it will have a dominant piece that looks like
\begin{align}
(\rho_L^a)_{\text{dom.}}  = \sum_{\tau',\tau} \left( \mathbb{P}_Q - \mathbb{P}_Q (\Lambda' + \Lambda'^\dagger) \mathbb{P}_Q + \cdots \right)_{\tau',\tau} \ket{\tau'}_L \bra{\tau}_L,
\label{eqn:rhoDom}
\end{align}
where $\cdots$ refers to higher order terms. Since in a suitable basis $\mathbb{P}_Q = \mathds{1}_a$ (for a fixed $\tilde{Q}$), the expression above can be approximated with an exponential ($1 - x \approx \exp(-x)$ for small $x$), and so the entanglement Hamiltonian is unitarily equivalent to $\mathbb{P}_Q (\Lambda' + \Lambda'^\dagger) \mathbb{P}_Q$. 

The original, flat, non-zero $2^{L_y/2- 1}$ eigenvalues of $\mathbb{P}_Q$ therefore become the non-zero $2^{L_y/2-1}$ eigenvalues of $\exp[-\mathbb{P}_Q(\Lambda' +  \Lambda'^\dagger) \mathbb{P}_Q]$, i.e.
\begin{align}
1 \to \text{eig}(\exp[-\mathbb{P}_Q(\Lambda' +  \Lambda'^\dagger) \mathbb{P}_Q]),
\end{align}
which are near $1$ in magnitude. The entanglement spectrum $\xi_{\text{ent.}}$ associated with these eigenvalues is then calculated by taking $-1$ times the logarithm of the eigenvalues of the reduced density matrix: $\xi_{\text{ent.}} = \text{eig}(\mathbb{P}_Q(\Lambda' + \Lambda'^\dagger) \mathbb{P}_Q) $. Of course, there will be other levels in the entanglement spectrum coming from the sub-dominant part of $\rho_L^a$, but they `flow down from infinity', as those $\xi_{\text{ent.}} \sim -\ln (\epsilon) \sim \infty $. These are not the levels of interest and will therefore be ignored. Thus, we aim to derive only the new eigenvalues of the reduced density matrix that are perturbed from the original {\it non-zero} values, up to {\it leading order} corrections. These values are defined to be the ones that give rise to the relevant part of the entanglement spectrum, and are the values of interest when comparing the edge to the ES in the edge-ES correspondence.

\begin{figure}\center
%\scalebox{0.65}{\input{HilbertSpace}}
\includegraphics[width=1.05\columnwidth]{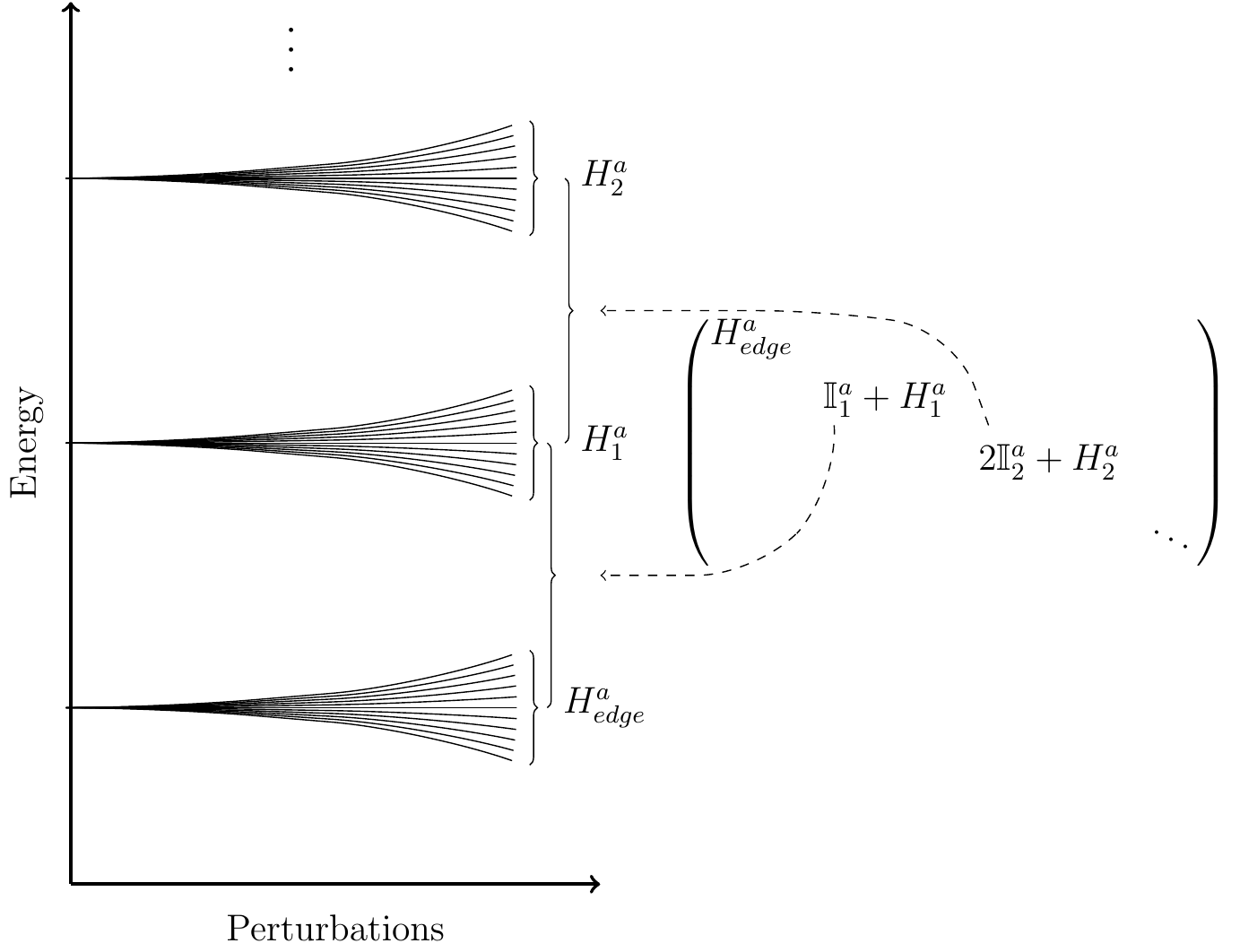}
\caption{  Illustration of the terms appearing in the first line of \eqn{eqn:Decomposition1}. The Schrieffer-Wolff transformation rotates the perturbed Hamiltonian into a Hamiltonian that is block diagonal in the old eigenspaces (so that $P_{\alpha}^a = \mathbb{I}_{\alpha}^a$), and adds small dynamics due to the perturbation on top of each space $(H_{\alpha}^a \simeq \mathcal{O}(\epsilon))$.}
\label{fig:Hilbert}
\end{figure}

Of course, the discussion above was to sketch the flow of the logic of our argument, and we have to be rigorous in our derivation. They key step to solve for the ground states of the perturbed system is to rewrite \eqn{eqn:pertH} in a way which makes obtaining the form of \eqn{eqn:genericState} manifest - in other words, we want to reorder the perturbative series which gives the tensor product structure automatically. The key is the Schrieffer-Wolff transformation. We can rewrite $H_L + \epsilon V_L$ within each topological sector (letting $a \simeq (\tilde{Q}_L, Q_L)$ also refer to the topological sector on the $L$ semi-infinite cylinder) as: 
\begin{align}
e^{-S^a_L} e^{S^a_L} P_{L}^a (H_L + \epsilon V_L) P_{L}^a e^{-S^a_L} e^{S^a_L},
\end{align}
where $P_{L}^a = \sum_{\alpha \geq 0} P_{\alpha,L}^a$, a projector onto the topological sector $a$. $P_{\alpha,L}^a$ are the projectors onto $\mathcal{V}_{\alpha,L}^a$ (see Sec.~\ref{sect:edgeExact}). This can be done since local perturbations do not connect topological sectors $a$ at any order in perturbation theory.

The BCH expansion then allows us to expand it as
\begin{align}
& e^{-S^a_L} ( H^a_{\text{edge}, L} + P^a_{1,L} + H^a_{1,L} + 2P^a_{2,L} + H^a_{2,L} + \cdots)e^{S^a_L} \nonumber \\
& = \underbrace{\sum_{\alpha \geq 1} \alpha P^a_{\alpha,L}}_{\text{large}} + \underbrace{ H^a_{\text{edge}, L} + \sum_{\alpha \geq 1} H^a_{\alpha,L} + \sum_{\alpha\geq1}[-S^a_L, \alpha P^a_{\alpha,L}]}_{\text{small}}  +\nonumber \\
& ~~~ \underbrace{ [-S^a_L,H^a_{\text{edge}, L}] + \sum_{\alpha \geq 1} [-S^a_L,H^a_{\alpha, L}] }_{\text{small}}  \nonumber \\
& ~~~  + \underbrace{\frac{1}{2} \sum_{\alpha \geq 1}[-S^a_L,[-S^a_L,\alpha P^a_{\alpha,L}]] + \cdots}_{\text{small}},
\label{eqn:Decomposition1}
\end{align}
and similarly for the theory on the $R$ semi-infinite cylinder, $H_R + \epsilon V_R$, using $a' \simeq (\tilde{Q}_R, Q_R)$ as the topological sector label. $H^a_{\alpha,L}$ are (at least) order $\epsilon$ effective Hamiltonians acting in $\mathcal{V}^a_{\alpha,L}$ that generate the dynamics over $\alpha P^a_{\alpha,L}$ upon the addition of perturbations $\epsilon V_L$ to the system. All terms labeled `large' are of order $1$, while all terms labeled `small' are at least of order $\epsilon$. To help illustrate the meaning of the terms that appear in the first line of the equation above between the exponentials, we have schematically plotted the energy spectrum of the Wen-plaquette model under perturbations in Fig.~\ref{fig:Hilbert}. The Schrieffer-Wolff transformation simply rotates the perturbed Hamiltonian into a Hamiltonian that is block diagonal in the old eigenspaces (so that $P_{\alpha}^a = \mathbb{I}_{\alpha}^a$), and adds small dynamics due to the perturbation on top of each space $(H_{\alpha}^a \simeq \mathcal{O}(\epsilon))$.

Now, we see that  $H_0 = H_{LR} + \sum_{\alpha \geq 1} \alpha ( \sum_a P^a_{\alpha,L} + \sum_{a' }P^{a'}_{\alpha,R} )$ is nothing but the original, unperturbed Wen-plaquette Hamiltonian on the full cylinder [c.f. \eqn{eqn:unpertH}] whose ground states are given by \eqn{eqn:ZOGS}, with $a = a' \equiv (\tilde{Q},Q)$ (hence justifying the use of a single label $a$). The small terms in \eqn{eqn:Decomposition1} (and similarly for $R$) can therefore be thought of as perturbations to the large Hamiltonian, $H_0$, and their corrections to the non-degenerate ground state (within each $a$ sector) can be calculated in normal non-degenerate wavefunction perturbation theory.

Similarly, we can decompose $\epsilon V_{LR}$ as
\begin{align}
\epsilon V_{LR} = \epsilon \sum_{\substack{\alpha \geq 0 \\ \tilde{Q}_L, \tilde{Q}_R}} (P_{\alpha,L}^{\tilde{Q}_L} \otimes P_{\alpha,L}^{\tilde{Q}_R}) V_{LR} (P_{\alpha,L}^{\tilde{Q}_L} \otimes P_{\alpha,L}^{\tilde{Q}_R})
\label{eqn:Decomposition2}
\end{align}
because $V_{LR}$ is local and cannot mix different $\tilde{Q}_L$ sectors (and similarly for $\tilde{Q}_R$).

This decomposition allows us to identify the origins of the corrections of $\Lambda, \Theta, \Xi$ and $\Omega$, at least to the first order process to the corrections to the ground states, \eqn{eqn:ZOGS}. For a fixed topological sector $a$, standard non-degenerate wavefunction perturbation theory tells us to correct the ground states in the first order process as
\begin{align}
\ket{\tilde{Q},Q} = \ket{\tilde{Q},Q}_0 + \sum_{\text{exc.}} \frac{\bra{\text{exc.}} O_{\text{small}} \ket{\tilde{Q},Q}_0 } {\Delta E} \ket{\text{exc.}},
\end{align}
where $O_{\text{small}}$ are small corrections to the unperturbed Hamiltonian, $\ket{\text{exc.}}$ excited eigenstates of the exact Wen-plaquette model, and $\Delta E$ the energy difference of the excited and ground states, which is always negative.

Thus we see that since $ H^a_{\text{edge},L} + H^a_{\text{edge},R} : \mathcal{V}^a_{0,L} \otimes \mathcal{V}^a_{0,R} \to \mathcal{V}^a_{0,L} \otimes \mathcal{V}^a_{0,R}$, it gives rise to part of the correction $\Lambda$. The other correction in $\Lambda$ comes from $(P_{0,L}^{\tilde{Q}} \otimes P_{0,R}^{\tilde{Q}} ) V_{LR} (P_{0,L}^{\tilde{Q}} \otimes P_{0,R}^{\tilde{Q}} )$. Furthermore, both terms which contribute are symmetric under the $\mathbb{Z}_2$ generators $\hat{Q}_L$ and $\hat{Q}_R$ and so do not mix the $Q_L$ and $Q_R$ topological sectors. Together, $\Lambda$ must be therefore $Q$ symmetric, i.e. $\Lambda = \mathbb{P}_Q \Lambda' \mathbb{P}_Q$, for some $\Lambda'$, as claimed. The exact derivation of $\Lambda'$ is left to Appendix \ref{sect:AppendixB}. It turns out that $\Lambda'$ is a Hermitian matrix, so that $\Lambda' = \Lambda'^\dagger$.
% Appendix

Next, since $S^a_L$ is block diagonal, $[-S^a_L, nP^a_{n,L}] : \mathcal{V}^a_{0,L} \otimes \mathcal{V}^a_{0,R} \to \mathcal{V}^a_{n,L} \otimes \mathcal{V}^a_{0,R}$, and generates $\Xi$ (the $R$ equivalent generates $\Theta$). Similarly, other terms in the decomposition of $V_{LR}$ also contribute to $\Theta, \Xi$ and $\Omega$.

Obviously, it is impossible to compute corrections to the ground state exactly for arbitrary local perturbations. However, it is still possible to say something concrete for generic arbitrary perturbations. In general, $\Lambda$ is of order $\epsilon$. We can  then ignore $\Theta$ and $\Xi$ as they give rise to $\epsilon^2$ corrections in the eigenvalues of the reduced density matrix. Since we are only concerned about corrections to the ground state which give the leading order corrections to the entanglement energies (which are order $\epsilon$ from $\Lambda$), we only need to keep corrections in the dominant part of the reduced density matrix. The perturbed wavefunction, considered to leading order, is then
\begin{align}
\ket{\tilde{Q},Q} = \frac{1}{\sqrt{Z}}\sum_{\tau',\tau} ( \mathbb{P}_Q - \mathbb{P}_Q \Lambda' \mathbb{P}_Q )_{\tau',\tau} \ket{\tau'}_L \ket{\tau}_R,
\end{align} 
$Z$ being a normalization factor. The reduced density matrix to leading order is then given by \eqn{eqn:rhoDom},
\begin{align}
\rho_L^a &\equiv \frac{1}{Z} \exp\left[-H_{\text{ent.}}^a\right] = \text{Tr}_R \ket{\tilde{Q}, Q}\bra{\tilde{Q}, Q} \nonumber \\
& \approx \frac{1}{Z} \left( \mathbb{P}_Q - 2\mathbb{P}_Q \Lambda' \mathbb{P}_Q \right) \nonumber \\
& \approx \frac{1}{Z} \exp\left[- 2 \mathbb{P}_Q \Lambda' \mathbb{P}_Q   \right],
\end{align}
leading to the identification
\begin{align}
H_{\text{ent.}}^a \equiv 2\mathbb{P}_Q \Lambda' \mathbb{P}_Q.
\label{eqn:Hent}
\end{align}
The entanglement spectrum, $\xi_{\text{ent.}}^a$, is given by
\begin{align}
(\xi_{\text{ent.}}^a)_n =  2\text{eig}( \mathbb{P}_Q \Lambda' \mathbb{P}_Q  )_n 
\label{eqn:pertES}
\end{align}
where $n = 1,2,\cdots,2^{L_y/2 - 1}$ label the eigenvalues of eigenvectors that have $\mathbb{P}_Q = 1$.

\vspace{5pt}
To conclude:
\vspace{5pt}

\noindent 
$\bullet$ 
The entanglement Hamiltonian, $H_{\text{ent.}}^a$, generically appearing at order $\epsilon$ ,  is given by \eqn{eqn:Hent}, where $\Lambda'$ is generated at lowest order by $H_{\text{edge},L}^a + H_{\text{edge},R}^a$ and $V_{LR}$ (see Appendix \ref{sect:AppendixB}). It is $\mathbb{Z}_2$ symmetric and acts on a virtual spin chain of $L_y/2$ sites, similar to the edge Hamiltonians.

\subsection{Edge-ES correspondence}
\label{sect:pertEdgeES}
From the calculation of $\Lambda'$ in Appendix \ref{sect:AppendixB} and \eqn{eqn:pertES}, we see that within each topological sector $a$, the edge Hamiltonian $H_{\text{edge},L}^a + H_{\text{edge},R}^a$ and the entanglement Hamiltonian  $H_{\text{ent.}}^a$ differ from each other in two ways: (i) terms in the edge Hamiltonian are reproduced in the entanglement Hamiltonian but with term-dependent rescaling factors, and (ii) additional terms arising from $V_{LR}$ appear in the entanglement Hamiltonian. Since this means that $H_{\text{edge},L}^a + H_{\text{edge},R}^a$ and $H_{\text{ent.}}^a$ can differ in a potentially arbitrary fashion, there is no reason to expect that the edge spectrum will match the entanglement spectrum, even for the low energy values. We therefore conclude that there is no edge-ES correspondence in general.

However, we note that both Hamiltonians have remarkably similar structure: they both act on a spin-$1/2$ chain of length $L_y/2$, and are $\mathbb{Z}_2$ invariant, i.e. they commute with the global spin flip operator $\prod_i \tau_i^x$, which is the Wilson loop operator in the bulk. The $\mathbb{Z}_2$ symmetry can therefore be understood as being enforced by the bulk topological order.

\vspace{5pt}
To summarize:
\vspace{5pt}

\noindent 
$\bullet$ 
There is no edge-ES correspondence, even in the low energy limit, for generic local perturbations. However, both the edge and entanglement Hamiltonians are $\mathbb{Z}_2$ symmetric and act on a virtual spin chain of $L_y/2$ sites.

\subsection{Remarks}
One might ask: what happens if $\Lambda$ appears at order $\epsilon^n$, $n \geq 2$ instead of at order $\epsilon$? Of course, this situation is a result of fine-tuning the perturbations to the system. For example, perturbing the Wen-plaquette model with only single-site magnetic fields (so that $V_{LR} = 0$) will yield an edge Hamiltonian at order $\epsilon^2$, and therefore $\Lambda$ at order $\epsilon^2$ as well. One has to now account for the additional contributions from $\Theta, \Xi$ in \eqn{eqn:genericState} as they will lead to $\epsilon^2$ corrections in the dominant part of the reduced density matrix, thereby potentially modifying the entanglement spectrum.

However, the procedure to account for these additional contributions is clear: we simply perform non-degenerate wavefunction perturbation theory on $\ket{\tilde{Q},Q}$ to a desired order consistently in the reduced density matrix, using the decomposition \eqn{eqn:Decomposition1} and \eqn{eqn:Decomposition2}. 

An explicit example showing how this is done is given in the next section (also refer to appendix \ref{sect:AppendixC}), where we consider a mechanism to achieve an edge-ES correspondence in the low energy limit. We look at the case of uniform single-site magnetic fields as perturbations and present the perturbative calculations to order $\epsilon^2$ explicitly. In that case, we will see that the terms $\Theta, \Xi$ simply lead to a constant shift in the entanglement energies of the entanglement spectrum, and so the relation that $H_{\text{ent.}}^a \equiv 2\mathbb{P}_Q \Lambda' \mathbb{P}_Q$ still holds up to a constant shift.

%%%%%%%%%%%%%%%%%%%%%%%%%%%%%%%%%%

\section{Mechanism for correspondence: Translational symmetry and Kramers-Wannier duality}
\label{sect:KW}
In this section, we present a mechanism that ensures an edge-ES correspondence, at least in a finite domain in Hamiltonian space. We consider the Wen-plaquette model as a symmetry enriched topological phase (SET) with the global symmetry being translational invariance along the edge/entanglement cut. That is, we restrict ourselves to perturbations that are translationally invariant along the width of the cylinder. In that case, both the edge and entanglement Hamiltonians will be Kramers-Wannier (KW) self-dual.

 It is not difficult to understand why this is so for the edge Hamiltonian. For the Wen-plaquette model and for an even $L_y$ cylinder, we have assigned a consistent checkerboard coloring of the plaquettes (see Fig.~\ref{fig:SemiInfiniteCylinder}), in which $e$ quasiparticles live on one color and $m$ quasiparticles live on the other color. However, this coloring is not unique, and we could have swapped the two colors, effectively exchanging $e \leftrightarrow m$ quasiparticles, a so-called electromagnetic duality. One can check that the fusion rules obeyed by the anyons are invariant under this swap. This swap can also be  thought of being effected by simplying translating the Wen-plaquette model by one site around the circumference of the cylinder, while keeping the underlying checkerboard coloring.

In terms of boundary operators, one sees that this swaps $S_{p,L}$ with $S_{\tilde{p},L}$, which necessarily leaves the physics of the edge or entanglement Hamiltonian invariant. However, recall that in the spin chain language $S_{p,L} \simeq \tau_{\tilde{p},L}^x$ with $S_{\tilde{p},L} \simeq \tau_{\tilde{p},L}^z \tau_{\tilde{p}-1,L}^z$. These two operators are precisely the Kramers-Wannier duals of each other. Translation by one site in the Wen-plaquette model thereby effects a KW transformation in the spin chain.

Thus, if we restrict to translationally invariant perturbations, then a term that appears in the edge Hamiltonian must also have its Kramers-Wannier dual appear in the edge Hamiltonian with the same coefficient, since $S_p \leftrightarrow S_{\tilde{p}}$ leaves the physics invariant. Thus, the edge Hamiltonian is Kramers-Wannier self-dual. It also follows that $\Lambda$, and therefore the entanglement Hamiltonian is also Kramers-Wannier self-dual, as claimed. 

Let us now analyze the edge and entanglement Hamiltonians in different $\tilde{Q}$ sectors, $H_{\text{edge}}^{\tilde{Q}}$ and $H_{\text{ent.}}^{\tilde{Q}} = 2 \Lambda'$, respectively. These are local, Kramers-Wannier self-dual, $\mathbb{Z}_2$ symmetric Hamiltonians on a finite length spin chain, with $\tilde{Q}$ giving the boundary conditions: $\tilde{Q} = +1$ corresponds to periodic boundary conditions and $\tilde{Q} = -1$ to anti-periodic boundary conditions. We claim that these Hamiltonians must be sitting at a phase transition.

First we argue the following. {\it Let us be given a local gapped $\mathbb{Z}_2$ symmetric spin Hamiltonian acting on an infinite spin chain. Let us assume that it spontaneously breaks the $\mathbb{Z}_2$ symmetry. Its Kramers-Wannier dual is another local gapped Hamiltonian which does not break spontaneously the dual $\mathbb{Z}_2$ symmetry. The reverse is true if the original Hamiltonian does not break the $\mathbb{Z}_2$ symmetry; then, its KW dual will spontaneously break the dual $\mathbb{Z}_2$ symmetry .}

The proof goes as follows. Consider the Hamiltonian on a finite spin chain. Since it is $\mathbb{Z}_2$ symmetric, it must be a function of $\tau_i^x$ and  $\tau_{i}^z \tau_{i+1}^z$ only. The Hilbert space associated with this spin chain be be split into the $\pm 1$ eigenvalues of the $\mathbb{Z}_2$ symmetry which is effected by the global spin-flip operator $\mathcal{S} \equiv \prod_i \tau_i^x$. We also have to assign boundary conditions, of which there are two kinds: periodic bondary conditions (PBC) and anti-periodic boundary conditions (APBC). Defining $\mathcal{T} \equiv \prod_i \tau_i^z \tau_{i+1}^z$ gives us $\mathcal{T} = + 1$ for PBC and $\mathcal{T} = -1$ for APBC. We can consider a mapping between this quantum system and another quantum system whose Hilbert space comprises of spins (labeled by $i+\frac{1}{2}$) placed on the links (labeled by $(i, i+1)$) of the original spin chain, i.e. a dual spin chain. Clearly, the two Hilbert spaces' dimensions are equal. Then, we can write down a dual Hamiltonian with the same spectrum as the original Hamiltonian, with the identification that the dual Hamiltonian is formed from the old Hamiltonian with $\tilde{\tau}_{i + \frac{1}{2}}^x \equiv  \tau_i^z \tau_{i+1}^z$ and $\tilde{\tau}_{i-\frac{1}{2}}^z \tilde{\tau}_{i+\frac{1}{2}}^z \equiv \tau_i^x$. This works because the new operators defined above give a representation of the algebra of the old set of operators. We see that this is nothing but the Kramers-Wannier transformation. 

However, note that the KW transformation maps $\mathcal{T} \leftrightarrow \tilde{\mathcal{S}}$ and $\mathcal{S} \leftrightarrow \tilde{\mathcal{T}}$, where $\tilde{\mathcal{S}}$ is the dual $\mathbb{Z}_2$ symmetry generator $ \prod_i \tilde{\tau}_i^x$ and $\tilde{\mathcal{T}}$ the dual boundary condition selector $\prod_i \tilde{\tau}_i^z \tilde{\tau}_{i+1}^z$. Now, if we assume that the original Hamiltonian with $\mathcal{T} = +1$ spontaneously breaks the $\mathbb{Z}_2$ symmetry $\mathcal{S}$, then it has two ground states that can be labeled by $\mathcal{S} = \pm 1$ which are close in energy. The order at which the ground state degeneracy is broken is at order $e^{-mL}$ for some mass scale $m$. Conversely, the Hamiltonian with $\mathcal{T} = -1$ will have a ground state with a domain wall between the above two vacua, and will thus have higher energy than the ground states on the system with $\mathcal{T} = +1$, with an energy difference on the order of the mass gap. On the other hand, for a $\mathbb{Z}_2$ preserving theory on $\mathcal{T} = \pm 1$, then there will only be a single ground state with $\mathcal{S} = +1$. The difference in energies of these ground states with $\mathcal{T} = \pm 1$ will be exponentially small.

Now, let us take the limit as the length of the chain becomes infinite. In this limit, the boundary conditions do not matter, and we  should only consider sectors with different $\mathcal{S}$ of the original Hilbert space and different $\tilde{\mathcal{S}}$ of the dual Hilbert space. From our exposition above, if the original Hamiltonian breaks the $\mathbb{Z}_2$ symmetry, then it will have two degenerate ground states labeled by $\mathcal{S} = \pm 1$ with some energy $E$. However, after the KW transformation, the dual Hamiltonian will now have only one state near $E$ (it has $\tilde{\mathcal{S}} = +1$) - all other states have higher energies, with an energy difference of at least the mass gap. Thus we see that the dual Hamiltonian does not break the dual $\mathbb{Z}_2$ symmetry. Similarly, if the original Hamiltonian does not break the $\mathbb{Z}_2$ symmetry, it has only one state with $\mathcal{S} = +1$ that has lowest energy $E'$; all other states are separated in energy by the mass gap. However, its dual Hamiltonian will have two states near energy $E'$, with $\tilde{\mathcal{S}} = \pm 1$. Thus, we see that the dual Hamiltonian spontaneously breaks the dual $\mathbb{Z}_2$ symmetry. This thereby concludes the proof of our claim.

Next, let us be given a Kramers-Wannier $\mathbb{Z}_2$ symmetric self-dual Hamiltonian $H_0$, and deform it by a small $\mathbb{Z}_2$ symmetric  perturbation $H_1$ which is odd under KW duality:
\begin{align}
H = H_0 + h H_1,
\end{align}
where a KW transformation maps $h \to -h$. If the deformed Hamiltonian $H$ is gapped and breaks the $\mathbb{Z}_2$ symmetric spontaneously for some sign of $h$, it will not break it for the opposite sign, and viceversa. Thus the theory must have a phase transition at $h = 0$! This therefore concludes the proof that $H_{\text{edge}}^{\tilde{Q}}$ and $H_{\text{ent.}}^{\tilde{Q}}$ must be sitting at a phase transition. 

Furthermore, it is natural to expect the phase transition to be generically of second order. If that was the case, the KW self-dual Hamiltonians would be gapless (i.e. critical). Now, a stronger statement can be made: if the dominant part of $H_{\text{edge}}^{\tilde{Q}}$ and $H_{\text{ent.}}^{\tilde{Q}}$ are the transverse field Ising model ($(1+1)$-d Ising model), then the low energy spectra of both Hamiltonians must be that of the $c = 1/2$ Ising CFT. This is guaranteed because there are no relevant deformations to the critical $(1+1)$-d Ising model that are both $\mathbb{Z}_2$ invariant and KW-even. To see this is true, list all relevant operators in the theory (i.e. with scaling dimension $\Delta < 2$): the spin primary field $\sigma$ and the energy density primary field $\epsilon$. Now $\sigma$ is $\mathbb{Z}_2$ even, so this deformation does not appear in the edge and entanglement Hamiltonians. On the other hand, $\epsilon$ is $\mathbb{Z}_2$ even but is KW-odd, so it cannot appear either. Any deformations in the edge and entanglement Hamiltonians must therefore be irrelevant - all renormalization flows are towards the $c = 1/2$ Ising CFT.

This therefore shows that there is a finite domain in Hamiltonian space such that the edge and entanglement Hamiltonians both realize the $c = 1/2$ Ising CFT as their low energy effective theory.  It is therefore seen that in this case, the Wen-plaquette model, considered as an SET with the global symmetry being translational invariance along the edge/entanglement cut, realizes an edge-ES correspondence: the low lying values of both the edge and entanglement spectra will match. 

However, a note of caution should be pointed out here. There is no guarantee that the edge and entanglement Hamiltonians will be both near the critical $(1+1)$-d Ising models, even though it is natural to assume they should be. There are other models which are also $\mathbb{Z}_2$ symmetric and Kramers-Wannier self-dual, such as the tricritical Ising model or even $\mathbb{Z}_2$ symmetric Hamiltonians which break the KW-duality spontaneously (i.e. the model realizes a first order transition, which the critical Ising model and tricritical Ising model do not).

Thus, one cannot conclude that the edge and entanglement Hamiltonians must  {\it both} be the $c = 1/2$ Ising CFT. For example, it could be that the edge Hamiltonian is a critical Ising model while the entanglement Hamiltonian is instead a tricritical Ising model. In that case, there is no edge-ES correspondence in the way that we have defined, as the low energy spectra clearly do not match. However, what is still guaranteed with the global translational symmetry is that both the edge and entanglement Hamiltonians will be $\mathbb{Z}_2$ symmetric and Kramers-Wannier dual; thus, a weaker form of edge-ES correspondence holds in which both Hamiltonians belong to the same {\it class} of Hamiltonians, whether or not the specific form of the Hamiltonian is the critical Ising, tricritical Ising, or a first order phase transition Hamiltonian.

\vspace{5pt}
In summary:
\vspace{5pt}

\noindent
$\bullet$ 
Considered as an SET, the symmetry protection (translational invariant perturbations) and the bulk topological order ensure that the edge and entanglement Hamiltonians of the Wen-plaquette model are Kramers-Wannier self-dual and are $\mathbb{Z}_2$ symmetric. If both theories are close to the critical $(1+1)$-d Ising model, then since there are no relevant KW-even and $\mathbb{Z}_2$ symmetric perturbations to the model, this implies that there is a finite domain in Hamiltonian space in which their low energy physics is a $c=1/2$ Ising CFT. There is, therefore, an edge-ES correspondence in such a scenario (the low energy spectra match). However, there is no guarantee that both Hamiltonians will always be critical Ising models, as there are other $\mathbb{Z}_2$ symmetric models which realize the Kramers-Wannier self-duality, such as the tricritical Ising model or a Hamiltonian sitting at a first order phase transition. Thus, a weaker form of the edge-ES correspondence instead holds, in which both the edge and entanglement Hamiltonians belong to the same class of Kramers-Wannier even and $\mathbb{Z}_2$ symmetric Hamiltonians.

{\color{red}}
\subsection{Analytical example: uniform single-site magnetic fields}
\label{sect:AExample}
We present an analytic and numerical illustration of our claim of the edge-ES correspondence, in which both the edge and entanglement Hamiltonians realize the critical $(1+1)$-d Ising models and thus have the $c = 1/2$ Ising CFT as their low energy effective theory. Let us consider the case of perturbations being uniform single-site magnetic fields:
\begin{align}
\epsilon V = \epsilon \sum_i h_X X_i + h_Y Y_i + h_Z Z_i,
\label{eqn:perturbation}
\end{align}
where $(h_X, h_Y, h_Z)$ is a vector with entries of order $1$. This perturbation is translationally invariant along the edge/cut of the cylinder. On physical grounds it is the simplest possible local perturbation of the toric code, and on theoretical grounds, it is the deformation of the Toric code that is the most well studied, see existing literature on the subject, e.g. \Ref{Tupitsyn, Yu, Vidal}.
% Cite deformation papers

Such a uniform single-site magnetic field generates an edge and entanglement Hamiltonian with the critical $(1+1)$-d Ising model as their dominant term  at order $\epsilon^2$: \eqn{eqn:edgeH} (Appendix \ref{sect:AppendixC_edge} gives the detailed calculations) tells us that the edge Hamiltonians of each topological sector are the critical periodic ($\tilde{Q} = +1$) /antiperiodic ($\tilde{Q} = -1$) transverse field Ising Hamiltonians projected into the $\Z_2$ sectors ($Q = \pm1$). Explicitly, the critical $(1+1)$-d Ising model on a spin chain of length $\tilde{L}_y/2$, also called the transverse field Ising model (TFIM), is given by
\begin{align}
H_{\text{TFIM}}^{\tilde{Q}} = -\sum_{\tilde{p}=1}^{\tilde{L}_y/2} \left( \tau_{\tilde{p}}^z \tau_{\tilde{p}-1}^z + \tau_{\tilde{p}}^x \right),
\label{eqn:TFIM}
\end{align}
with toroidal boundary conditions $\tau_{0} = \tilde{Q} \tau_{\tilde{L}_y/2}$, and its decomposition into its $\mathbb{Z}_2$ charge sectors as follows:
\begin{align}
H_{\text{TFIM},Q}^{\tilde{Q}} \equiv \mathbb{P}_{Q} H_{\text{TFIM}}^{\tilde{Q}} \mathbb{P}_{Q},
\end{align}
where $\mathbb{P}_{Q} = \prod_{\tilde{p}=1}^{\tilde{L}_y/2} \tau_{\tilde{p}}^x$. This model is the {\it lattice realization} of the $c = 1/2$ Ising CFT.

On the other hand, for the entanglement Hamiltonians (refer to Appendix \ref{sect:AppendixC_ent} for the calculation of the entanglement Hamiltonians), up to rescaling and shifting, we have the following identification between states in the topological phase (left) and their entanglement Hamiltonians to leading order ($\epsilon^2$) (right):
\begin{align}
\ket{\mathbb{I}} &\leftrightarrow H_{\text{TFIM},+1}^{+1}, \nonumber \\
\ket{e} &\leftrightarrow H_{\text{TFIM},-1}^{+1}, \nonumber \\
\ket{m} &\leftrightarrow H_{\text{TFIM},+1}^{-1}, \nonumber \\
\ket{\varepsilon} &\leftrightarrow H_{\text{TFIM},-1}^{-1},
\label{eqn:Prediction}
\end{align}
where the label $\{ \mathbb{I}, e, m, \varepsilon = e \times m\}$ indicates that the states carry the corresponding anyonic flux.

Thus, we see that there is an edge-ES correspondence in this case:  both the edge and entanglement Hamiltonians have the $c=1/2$ Ising CFT as their low energy effective theory.

\subsection{Numerical example: uniform single-site magnetic fields}
\label{sect:NExample}

\begin{figure*}\center
\includegraphics[width=1.55\columnwidth]{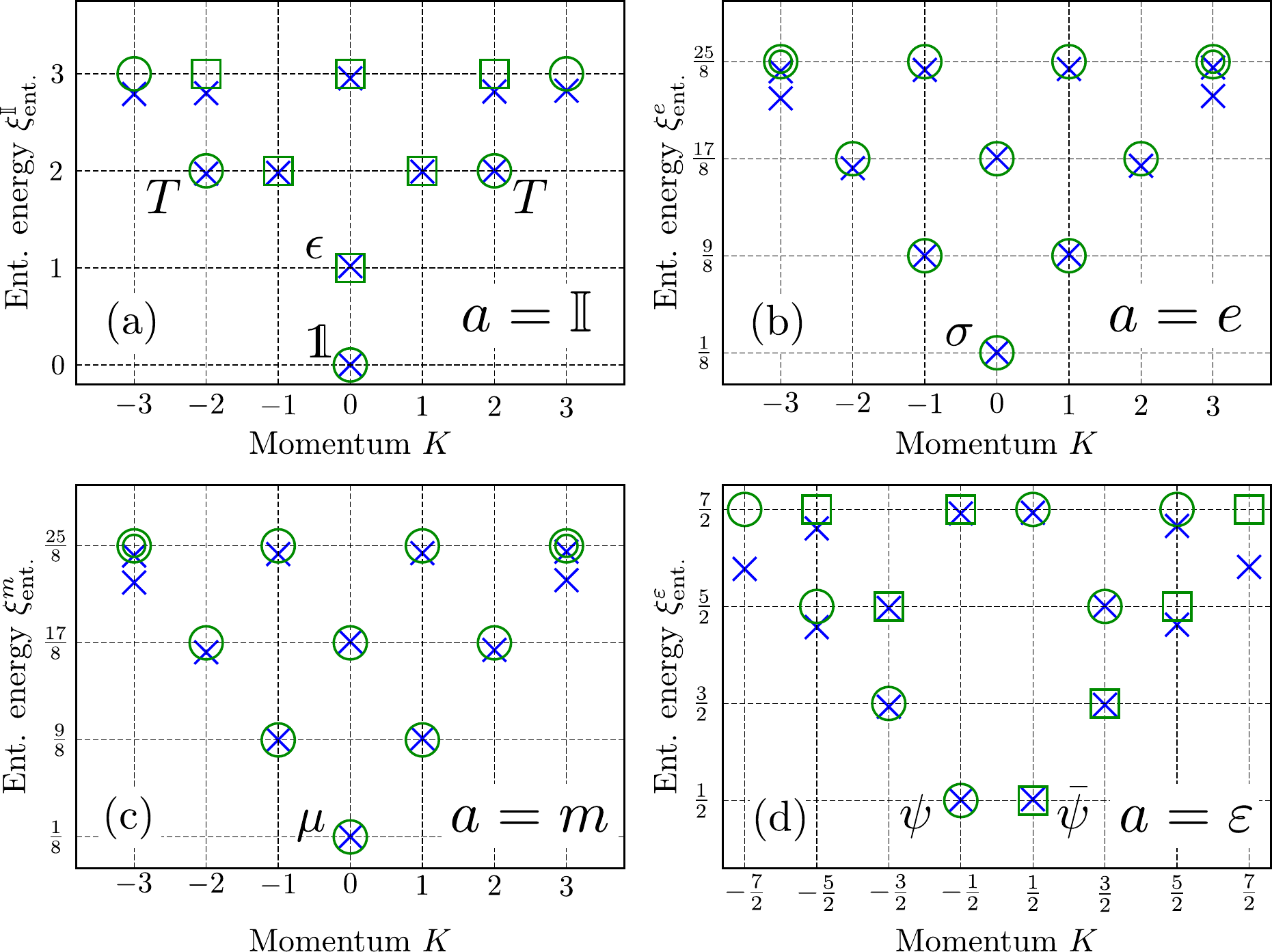}
\caption{(Color online). Rescaled and shifted entanglement spectra in each topological sector $a$ for Wen-plaquette model on an $L_y  = 20$ (even), infinite cylinder, for $\epsilon h_X = -0.01$, $\epsilon h_Y = -0.015$ and $\epsilon h_Z = -0.02$. The data points are marked by blue crosses. We have also plotted the $c = 1/2$ Ising CFT spectra in each sector for comparison (scaling dim.~versus momenta), with the identification that entanglement energy equals scaling dimension. They are given by the green squares and circles. Squares and circles represent conformal towers labeled by different primaries at the bottom of the towers, and double circles denote two-fold degenerate states. (a) The ES of the ground state $\ket{\mathbb{I}}$ with the identity anyonic flux through the cylinder. This corresponds to the $(\tilde{Q},Q) = (+1,+1)$ sector of the TFIM, whose low energy effective theory is the $c=1/2$ Ising CFT in the sector that the identity primary $\mathds{1}$ \& its descendants (green circles), and energy density primary $\epsilon$ \& its descendants (green squares) belong to. The two points labeled $T$ are the holomorphic and antiholomorphic stress-energy tensors. (b) ES of $\ket{e}$, the state with the electric anyonic flux.  This gives the $(\tilde{Q},Q) = (+1,-1)$ sector of the TFIM. This corresponds to the sector of Ising CFT which the spin primary $\sigma$ and its descendants belong to (green circles). (c) ES of $\ket{m}$, the state with the magnetic anyonic flux.  This gives the $(\tilde{Q},Q) = (-1,+1)$ sector of the TFIM, which in turn corresponds to the sector of the Ising CFT (with a $D_{\epsilon}$ defect insertion\cite{PetkovaZuber2001,Frohlich, Chui}) which the disorder primary $\mu$ and its descendants belong to (green circles). The spectra of the conformal families from $\sigma$ and $\mu$ conincide, and so do the entanglement spectra. (d) ES of $\ket{\varepsilon}$, the state with the fermionic anyonic flux.  This gives the $(\tilde{Q},Q) = (-1,-1)$ sector of the TFIM. This corresponds to the sector of the Ising CFT (with a $D_{\epsilon}$ defect insertion) which the two Majorana fermion primaries $\psi, \bar{\psi}$ and their descendants belong to (green squares and circles). }\label{fig:plotPBC}
\end{figure*}

To check our predictions, we numerically solve for the ground states of the Wen-plaquette model on an infinite cylinder with even $L_y$ sites on its circumference, $L_y = 20$, with perturbations of the form \eqn{eqn:perturbation}, using DMRG for infinite cylinders\cite{McCulloch, CincioVidal2012}. Here, $\epsilon h_X = -0.01$, $\epsilon h_Y = -0.015$ and $\epsilon h_Z = -0.02$. Starting from a random initialization of the MPS, it was found that the DMRG algorithm converged to four orthonormal states, which are the four ground states with well-defined anyonic flux through the cylinder. We then find the entanglement spectrum associated with each ground state, and plot it against momenta around the cylinder. Figure \ref{fig:plotPBC} shows the entanglement spectra in each topological sector $a$ of the Wen-plaquette model with the parameters described above.

\begin{figure*}\center
\includegraphics[width=1.55\columnwidth]{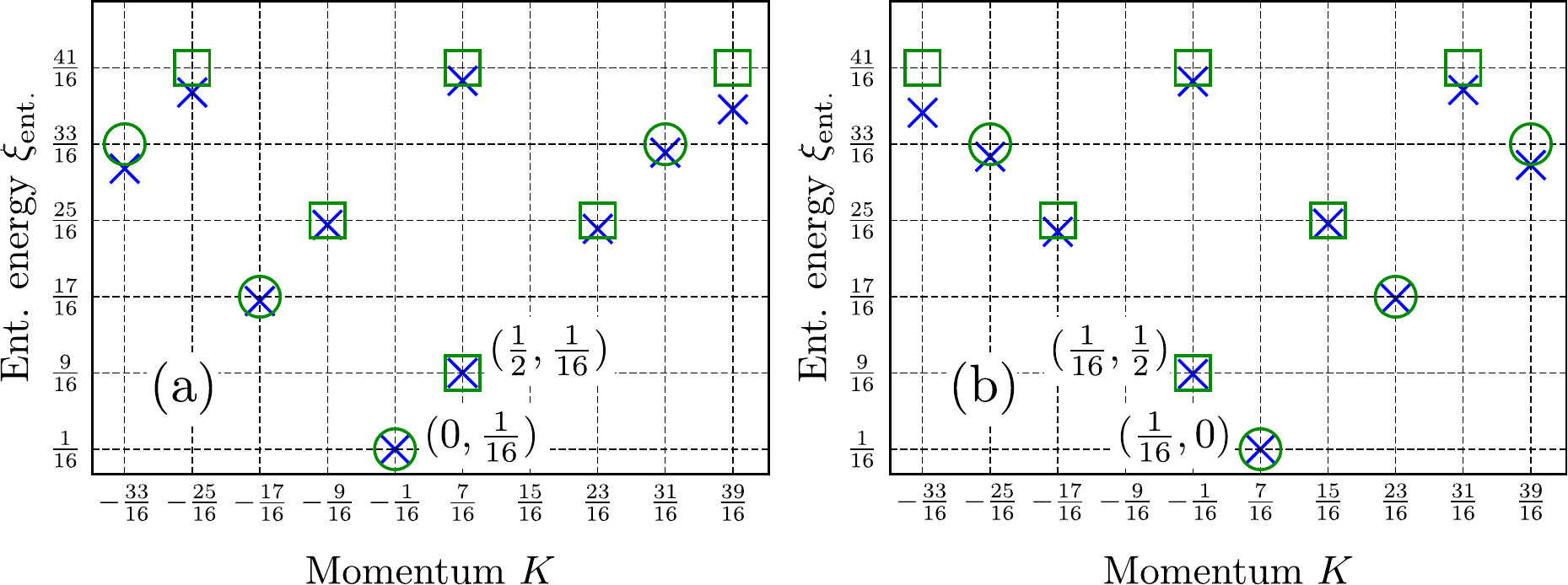}
\caption{(Color online). Rescaled and shifted entanglement spectra in each topological sector for the perturbed Wen-plaquette model on an $L_y = 19$ (odd), infinite cylinder. Shown also are the $c  = 1/2$ CFT spectra with a conformal duality defect $D_\sigma$ (see Fig.~\ref{fig:plotPBC} for details on labels). (a) The ES corresponds to the spectrum of the duality-twisted Ising model with primary fields $(0,1/16),(1/2, 1/16)$. (b) The ES corresponds to the spectrum of the duality-twisted Ising model with primary fields $(1/16,0), (1/16,1/2)$. }\label{fig:plotAPBC}
\end{figure*}

Each spectrum has been (i) shifted and then (ii) rescaled by the same value across all topological sectors $a$. The shift has been chosen so that the lowest entanglement energy across all entanglement spectra is at entanglement energy $0$, and the rescaling chosen so that in the topological sector which contains the lowest overall entanglement energy, the lowest entanglement energy at momentum $K = \pm 2$ is at entanglement energy $2$. In doing so, we fix the identity primary (scaling dimension $\Delta = 0$, momentum $K = 0$) and the holomorphic and antiholomorphic stress-energy tensors (scaling dimension $\Delta = 2$, momentum $K = \pm2$), which are always present in a CFT without defects. Remarkably, this common shift and rescaling of the entanglement spectra across topological sectors $a$ reproduces the $c = 1/2$ Ising CFT spectra in the different charge sectors accurately; from Fig.~\ref{fig:plotPBC}(a), we can identify the identity primary $\mathds{1}$ [conformal weights $(h,\bar{h}) = (0,0)$] and the energy-density primary $\epsilon\sim(1/2,1/2)$ with its descendants, which belong to the $(\tilde{Q},Q) = (+1,+1)$ charge sector of the TFIM. In the CFT language this is the $\mathbb{Z}_2 = +1$ sector of the usual diagonal $c = 1/2$ Ising CFT. From Fig.~\ref{fig:plotPBC}(b), we identify the spin primary $\sigma\sim(1/16,1/16)$ with its descendants, which belong to the $(\tilde{Q},Q) = (+1,-1)$ sector of the TFIM. This corresponds to the $\mathbb{Z}_2 = -1$ sector of the Ising CFT. From Fig.~\ref{fig:plotPBC}(c), we identify the disorder primary $\mu\sim(1/16, 1/16)$ with its descendants, which is identical to the spectrum of $\sigma$, belonging to the $(\tilde{Q},Q) = (-1,+1)$ sector of the TFIM. In this case, this corresponds to the $\mathbb{Z}_2 = +1$ sector of $c = 1/2$ Ising CFT with anti-periodic boundary conditions. Lastly, we identify the Majorana fermions $\psi\sim(1/2,0)$ and $\bar{\psi}\sim(0,1/2)$ and its descendants which belong to the $(\tilde{Q},Q) = (-1,-1)$ sector of the TFIM. This corresponds to the $\mathbb{Z}=-1$ sector of the Ising CFT with anti-periodic boundary conditions. This is in agreement with the theoretical prediction from our calculation, \eqn{eqn:Prediction}.

Lastly, even though our analysis in this paper was restricted to the Wen-plaquette model on an infinite cylinder with even $L_y$ circumference, we numerically solve for the ground states of the perturbed Wen-plaquette model on an infinite cylinder with odd $L_y$ circumference.  The perturbation is as before, \eqn{eqn:perturbation}, with values as in the even cylinder case. We take $L_y = 19$. In this case, a consistent checkerboard coloring of the plaquettes cannot be done. If one insists on placing a checkerboard coloring on the cylinder, there is necessarily a line of topological defects where an $e$ quasiparticle (as measured locally) transmutes into an $m$ quasiparticle upon crossing the line defect\cite{Bombin}. 
% Cite oliver? Hector?

In this case, we obtain two ground states. This makes sense as the only Wilson loop operator that exists is a string that wraps around the cylinder twice. The two ground states can be taken to be eigenvectors of this Wilson loop operator.  Figure \ref{fig:plotAPBC} shows the plots of the entanglement energies against momenta along a $L_y = 19$ cylinder for both ground states, with a common shifting and rescaling as follows. We shift the spectra of the two ground states so that the lowest entanglement energy of one sector is at $1/16$, and we rescale both ES so that the next lowest entanglement energy in that sector is at $9/16$. The two ES exactly match the spectrum of the boundary theory computed directly using perturbation theory
\begin{align}
H_{\pm} = - \left( \sum_{i=1}^{N-1} \tau_i^x  + \sum_{i=1}^{N-1} \tau_{i}^z \tau_{i+1}^z \pm \tau_{N}^y \tau_1^z \right),
\label{eqn:DualityTwistedH}
\end{align}
with $N = \frac{L_y + 1}{2}$, which is the Ising model with {\it duality-twisted} boundary conditions\cite{Grimm2002}. We see that $H_+$ and $H_-$ are related by complex conjugation, thus they have the same energy spectrum but with opposite momenta which can also be seen from the numerical data. Figure \ref{fig:plotAPBC}(a) shows the ES which corresponds to the spectrum of the duality-twisted Ising model with primary fields $(0,1/16),(1/2, 1/16)$, while figure \ref{fig:plotAPBC}(b) shows the ES which corresponds to the spectrum of the duality-twisted Ising model with primary fields $(1/16,0), (1/16,1/2)$.

All these results can be understood from a CFT point-of-view, using the language of conformal defects. The  $c=1/2$ Ising CFT can be given different twisted boundary conditions by insertions of conformal defect lines $X$, where the spectrum is then given by the \textit{generalized twisted partition function}\cite{PetkovaZuber2001} $Z_X = \tr\left(X\,q^{L_0-c/24}q^{\bar{L}_0-c/24}\right)$. In order to be able to move the defects around without energy cost, we need $X$ to commute with the energy-momentum tensor, or equivalently $[L_n,X]=[\bar L_n,X]=0$ where $L_n$ and $\bar L_n$ are the Virasoro algebra generators. The conformal defects are then classified by representations of the Virasoro algebra, for the $c=1/2$ CFT we have three defects: $D_{\mathds{1}}$, $D_{\epsilon}$ and $D_{\sigma}$. Crossing the $D_{\epsilon}$ defect $\sigma$ goes to $-\sigma$ and it thus implement anti-periodic boundary conditions, while for $D_{\sigma}$ we have $\sigma\rightarrow\mu$ which is nothing but the Kramers-Wannier duality. The spectra with these defect insertions are given by\cite{PetkovaZuber2001,Frohlich, Chui}
\begin{align*}
	Z_{\mathds 1} &= |\chi_0|^2+|\chi_{\frac 12}|^2+|\chi_{\frac 1{16}}|^2,\\
	Z_{\epsilon} &= |\chi_{\frac 1{16}}|^2+\chi_0\bar{\chi}_{\frac 12}+\chi_{\frac 12}\bar{\chi}_0,\\
	Z_{\sigma} &= (\chi_0+\chi_{\frac 12})\bar{\chi}_{\frac 1{16}}+\chi_{\frac 1{16}}(\bar{\chi}_0+\bar{\chi}_{\frac 12}).
\end{align*}
The spectra in Fig.~\ref{fig:plotPBC}(a) and \ref{fig:plotPBC}(b) correspond to the insertion of the trivial (identity) conformal defect $D_{\mathds{1}}$ in the partition function, while the spectra in Figs.~\ref{fig:plotPBC}(c)~and \ref{fig:plotPBC}(d) correspond to the insertion of the conformal epsilon defect $D_{\epsilon}$. These two cases are distinguished by the absence and presence of a Wilson flux line in the ground states of the bulk respectively, as measured by \scalebox{.55}{\GammaRed}. This Wilson flux line terminates at the boundary and is responsible for implementing anti-periodic BC, and thus can be thought of as `inserting' the conformal defect $D_{\epsilon}$ which has the same function. Next, the spectra in Fig.~\ref{fig:plotAPBC} correspond to the insertion of the conformal duality defect $D_{\sigma}$ in the partition function. This duality defect is nothing but the endpoint of the topological line defect in the bulk which comes from electro-magnetic duality. Thus this is completely consistent with the following correspondence between bulk and boundary: \textit{electro-magnetic duality $\leftrightarrow$ Kramers-Wannier duality}, as seen several times in this paper.
Perhaps this hints that one can understand defects in a CFT as arising from defects in a topological phase of one higher dimension, which has the CFT as its boundary theory\cite{RCFT}? 

Lastly, if one can find a mechanism to protect the edge-ES correspondence in such cases, then one can construct lattice realizations of CFTs with defects, by constructing the low-energy effective edge Hamiltonian using the Schrieffer-Wolff transformation  and identifying a mapping from boundary operators acting in the bulk to local operators acting on some effective low-energy degrees of freedom. In particular, for the case of the Wen-plaquette model on the odd cylinder, we find by direct perturbative calculations the lattice realization of a duality-twisted Ising model, which was only written down very recently by \Ref{Grimm2002}, \eqn{eqn:DualityTwistedH}.
%Grimm.

\section{Conclusion and discussion}
\label{sect:Conclusion}
In this paper, we have studied the edge-ES correspondence in a non-chiral topological phase, specifically concentrating on a phase with $\Z_2$ topological order, the Wen-plaquette model on an infinite cylinder. Our main result of the paper is a detailed, microscopic calculation of both the edge and entanglement Hamiltonians, which exhibits the mechanism for the absence or presence of the correspondence. We find through our calculation that the correspondence, i.e., that the edge and entanglement spectra agree in the low energy limit up to rescaling and shifting, is exact for the unperturbed Wen-plaquette model. However, for generic local perturbations, there is no edge-ES correspondence. 

We have managed to identify a mechanism to establish the edge-ES correspondence though, by considering the Wen-plaquette model as a SET with the global symmetry being translational invariance along the edge/entanglement cut. That is, if we deform the Wen-plaquette model by perturbations that are restricted to be invariant under translation by one site along the edge/cut, then both the edge and entanglement Hamiltonians are Kramers-Wannier self-dual. There is then a finite domain in Hamiltonian space such that both the edge and entanglement Hamiltonians have the $c = 1/2$ Ising CFT as their low energy effective theory, as there are no KW-even and $\mathbb{Z}_2$ symmetric perturbations to the critical $(1+1)$-d Ising model that are relevant. Thus, the edge-ES correspondence is achieved in such a scenario. However, there is no guarantee that both the edge and entanglement Hamiltonians will be near the critical Ising models, as there are other models which are Kramers-Wannier even and $\mathbb{Z}_2$ symmetric as well. Thus, a weaker form of the edge-ES correspondence holds, in which both Hamiltonians belong to the same class of Hamiltonians, even though the low energy spectra do not match.

\begin{comment}
We have achieved this by showing that for a bipartition of the infinite cylinder into two semi-infinite cylinders, the entanglement Hamiltonian $H_{\text{ent.}}^a$ of each ground state with well-defined anyonic flux is directly proportional to the sum of the corresponding physical edge Hamiltonians on the two semi-infinite cylinders. This result offers the first rigorous, microscopic proof of the correspondence in a non-chiral topological phase using perturbation theory, in contrast to other analytic proofs which rely on more advanced mathematical techniques that only apply to chiral topological phases, thereby extending our understanding of the edge-ES correspondence to more general topological phases. With our result, we can also study the edge physics of the $\Z_2$ spin liquid by considering its entanglement spectrum.
\end{comment}

Our approach of deriving an effective spin ladder Hamiltonian spanning the entanglement cut can potentially be extended to prove some form of the edge-ES correspondence in other non-chiral topological phases that are defined by a fixed point Hamiltonian consisting of a sum of mutually commuting local terms. One would have to find a set of maximally commuting operators on one half of the system, and also find a representation of the algebra of the operators acting on the edge to derive an analogous spin ladder system. In particular, we note that the extension of our calculations to the direct generalization of $\Z_2$ topological phases, $\Z_N$ topological phases, is straightforward, and we leave it to future work to explore the edge-ES correspondence in those cases.

\begin{acknowledgments}
We would like to thank Janet Hung for helpful discussions which led to this work. Research at Perimeter Institute is supported by the Government of Canada through Industry Canada and by the Province of Ontario through the Ministry of Research and Innovation, and also through the Ministry of Economic Development and Innovation. W.W.H.~and G.V.~acknowledge support by NSERC (discovery grant). L.C.~and G.V.~acknowledge support from the Templeton Foundation. G.V.~acknowledges support from the Simons Foundation. G.V.~thanks the Australian Research Council Centre of Excellence for Engineered Quantum Systems.
\end{acknowledgments}

\appendix

\section{Schrieffer-Wolff transformation}
\label{sect:AppendixA}
In this appendix we introduce the Schrieffer-Wolff transformation which reproduces the work of \Ref{BravyiDiVincenzoLoss}.

Let $H_0$ be a Hamiltonian that has a low-energy eigenspace $\mathcal{V}_0$ and a high-energy eigenspace $\mathcal{V}_1$ separated by a spectral gap. $H_0$ can be written as
\begin{align}
H_0 = P_0 H_0 P_0 + P_1 H_0 P_1,
\end{align}
where $P_\alpha$ are projectors onto $\mathcal{V}_{\alpha}$, where $\alpha = 0,1$. Let us add a small perturbation $\epsilon V$ which does not commute with $H_0$. Assuming the perturbation is weak enough, the new Hamiltonian will still have low and high-energy eigenspaces, $\tilde{\mathcal{V}}_0$ and $\tilde{\mathcal{V}}_1$, which have the same dimension as $\mathcal{V}_0$ and $\mathcal{V}_1$ respectively. That is, 
\begin{align}
H = H_0 + \epsilon V = \tilde{P}_0 H \tilde{P}_0 + \tilde{P}_1 H \tilde{P}_1.
\end{align}
Then there exists a unique direct rotation (i.e. unitary) $U$ that rotates the old and new subspaces
\begin{align}
U \tilde{P}_\alpha U^\dagger = P_\alpha.
\end{align}
By direct rotation, we mean the ``minimal'' rotation that maps $\tilde{P}_\alpha$ to $P_\alpha$: among all unitary operators $U'$ satisfying $U' \tilde{P}_\alpha U'^\dagger = P_\alpha$, the direction rotation $U$ differs least from the identity in the Frobenius norm. See \Ref{BravyiDiVincenzoLoss} for the construction of a direct rotation. In that case, we can rotate $H$ to a new Hamiltonian $H'$ with eigenspaces $\mathcal{V}_\alpha$:
\begin{align}
H' := U H U^\dagger = P_0 U H U^\dagger P_0 + P_1 U H U^\dagger P_1.
\end{align}
This is the so-called Schrieffer-Wolff transformation.

There exists a unique antihermitian and block-off diagonal (in both $\mathcal{V}_\alpha$ and $\tilde{\mathcal{V}}_\alpha$) operator $S$ such that $U = e^S$ and $||S|| < \pi/2$. It is constructed perturbatively as follows.

First we introduce some notation. Decompose an operator $X$ on the Hilbert space in its block-diagonal $X_{\text{d}}$ and block off-diagonal $X_{\text{od}}$ parts:
\begin{align}
&X_{\text{d}} = P_0 X P_0 + (1 - P_0) X (1 - P_0), \nonumber \\
&X_{\text{od}} = P_0 X (1- P_0) + (1-P_0) X P_0.
\end{align}
Also given $Y$, an operator acting on the Hilbert space, define the linear map which is the adjoint action of $Y$ on other operators $X$ that act on the Hilbert space:
\begin{align}
\hat{Y}(X) = [Y,X].
\end{align}
Lastly, define
\begin{align}
\mathcal{L}(X) = \sum_{i,j} \frac{\bra{i} X_{\text{od}} \ket{j} }{E_i - E_j} \ket{i}\bra{j},
\end{align}
where $\ket{i}$ is an eigenvector of $H_0$ with eigenvalue $E_i$ and similarly for $j$. Note that $\mathcal{L}(X)$ is by construction block off-diagonal. 

Now, $S$ can be written perturbatively as
\begin{align}
S = \sum_{n = 1}^\infty \epsilon^n S_n, ~~~ S_n^\dagger = -S_n,
\end{align}
where 
\begin{align}
& S_1 = \mathcal{L}(V_{\text{od}}), \nonumber \\
& S_2 = - \mathcal{L}\hat{V}_{\text{d}}(S_1), \nonumber \\
& S_n = - \mathcal{L}\hat{V}_{\text{d}}(S_{n-1}) + \sum_{j \geq 1} a_{2j} \mathcal{L} \hat{S}^{2j}(V_{\text{od}})_{n-1},  \text{ for } n \geq 3.
\end{align}
Here the coefficients $a_{m}$ come from the Taylor series
\begin{align}
x \coth(x) = \sum_{n= 0}^\infty a_{2n} x^{2n}, ~~~ a_m = \frac{2^m B_m}{m!},
\end{align}
where $B_m$ are the Bernoulli numbers. We have also used the shorthand
\begin{align}
\hat{S}^k(V_{\text{od}})_m = \sum_{\stackrel{n_1, \cdots, n_k \geq 1}{n_1 + \cdots + n_k = m}} \hat{S}_{n_1} \cdots \hat{S}_{n_k}(V_{\text{od}}).
\end{align}
From here, the low-energy effective Hamiltonian of $H'$ is given by
\begin{align}
H_{\text{eff}} := P_0 H' P_0 = P_0 H_0 P_0 + \epsilon P_0 V P_0 + \sum_{n=2}^\infty \epsilon^n H_{\text{eff},n},
\label{eqn:HeffAppendixSW}
\end{align}
where
\begin{align}
H_{\text{eff},n} = \sum_{j \geq 1} b_{2j - 1} P_0 \hat{S}^{2j - 1}(V_{\text{od}})_{n-1} P_0.
\end{align}
$b_{2j-1}$ are the Taylor coefficients of $\tanh(x/2)$, i.e.
\begin{align}
\tanh(x/2) = \sum_{n=1}^\infty b_{2n-1} x^{2n-1}, ~~~ b_{2n-1} = \frac{2(2^{2n} - 1) B_{2n} }{(2n)!}.
\end{align}
To second order in perturbation theory, and with a basis of states $\ket{i,\alpha} \in \mathcal{V}_\alpha$, the matrix elements of $H'$ projected into $\mathcal{V}_\alpha$ is given explicitly
\begin{align}
&\bra{i,\alpha}H'\ket{j,\alpha} = E^\alpha_i \delta_{ij} + \epsilon \bra{i,\alpha}V\ket{j,\alpha}   + \frac{\epsilon^2}{2} \sum_{\substack{k \\ \beta \neq \alpha} } \nonumber \\
& \bra{i,\alpha}V\ket{k,\beta} \bra{k, \beta}V\ket{j,\alpha}\left( \frac{1}{E_i^\alpha - E_k^\beta} + \frac{1}{E_j^\alpha - E_k^\beta} \right),
\label{eqn:EffHMethodAppendix}
\end{align}
where $E_{i}^\alpha$ is the energy of $\ket{i,\alpha}$. 

The generalized Schrieffer-Wolff transformation to Hamiltonians that have many invariant subspaces, each separated by a spectral gap to the next subspace, such that the Hilbert space $\mathcal{H} = \bigoplus_{\alpha \geq 0} V_{\alpha}$, is given in \Ref{Zheng}. The generator of rotation $S$ is still a block off-diagonal hermitian operator, with its first term in its expansion given by
\begin{align}
S_1 = \sum_{\substack{i, j, \alpha, \beta \\\alpha\neq\beta }} \frac{\bra{i,\alpha} V_{\text{od}} \ket{j,\beta}}{E_i^\alpha - E_j^\beta} \ket{i,\alpha}\bra{j,\beta},
\label{eqn:S1}
\end{align}
where $\ket{i,\alpha} \in \mathcal{V}_\alpha$, $\ket{j,\beta} \in \mathcal{V}_\beta$ and $E_{i}^\alpha$ is the energy of $\ket{i,\alpha}$. We will use this term in the derivation of the entanglement Hamiltonian. 
%http://journals.aps.org/prb/pdf/10.1103/PhysRevB.63.144410

\section{Calculation of $\Lambda'$ of the entanglement spectrum}
\label{sect:AppendixB}
In this appendix, we calculate $\Lambda'$, [see Sec.~\ref{sect:pertES} \eqn{eqn:genericState}], obtained from standard non-degenerate wavefunction perturbation theory. Our starting point is the perturbed Wen-plaquette Hamiltonian on the infinite cylinder, $H = H_L + H_R + H_{LR} + \epsilon(V_L + V_R + V_{LR})$, written according to the decompositions given by \eqn{eqn:Decomposition1} and \eqn{eqn:Decomposition2} using the Schrieffer-Wolff transformation. This reorganizes the perturbative series to make the tensor product structure of the new ground state manifest. We have:
\begin{align}
&H = \underbrace{H_{LR} + \sum_{\alpha \geq 1} \alpha \big( \sum_a P^a_{\alpha,L} + \sum_{a'} P^{a'}_{\alpha,R} \big)}_{\text{large}} \nonumber \\
& + \sum_a \bigg(  H^a_{\text{edge}, L} + [-S^a_L,H^a_{\text{edge}, L}] + \cdots + \sum_{\alpha \geq 1} \big( H^a_{\alpha,L} +  \nonumber \\
&   [-S^a_L, \alpha P^a_{\alpha,L}]  + [-S^a_L,H^a_{\alpha, L}]   +  \frac{1}{2}[-S^a_L,[-S^a_L,\alpha P^a_{\alpha,L}]] + \cdots \big) \bigg) \nonumber\\
& + \sum_{a'} \bigg( \cdots R \text{ terms} \cdots \bigg) \nonumber \\
& + \epsilon \sum_{\substack{\alpha \geq 0 \\ \tilde{Q}_L, \tilde{Q}_R}} (P_{\alpha,L}^{\tilde{Q}_L} \otimes P_{\alpha,L}^{\tilde{Q}_R}) V_{LR} (P_{\alpha,L}^{\tilde{Q}_L} \otimes P_{\alpha,L}^{\tilde{Q}_R}),
\label{eqn:fullHAppendix}
\end{align}
where the large part is simply the unperturbed Wen-plaquette model $H_L + H_R + H_{LR}$, and the other part is small (order $\epsilon$ or higher) which comes from the perturbations $\epsilon(V_L + V_R + V_{LR})$.

From section \ref{sect:unpertES}, it suffices to solve $H_{LR}$ for the four ground states of the Wen-plaquette model. The projection of $H_{LR}$ onto a $\tilde{Q}$ sector, under the mapping given by \eqn{eqn:Map}, becomes a spin-ladder Hamiltonian:
\begin{align}
H_{\text{eff}} = - \sum_{\tilde{p} = 1}^{\tilde L_y/2} (\tau_{\tilde{p},L}^{x} \tau_{\tilde{p},R}^{x} +  \tau_{\tilde{p},L}^{z} \tau_{ \tilde{p}-1,L}^{z} \tau_{\tilde{p},R}^{z} \tau_{\tilde{p}-1,R}^{z}),
\label{eqn:HeffAppendix}
\end{align}
whose ground states (in each topological sector $a \simeq (\tilde{Q},Q)$ ) are given by \eqn{eqn:ZOGS}:
\begin{align}
\ket{\tilde{Q},Q} = \frac{1}{\sqrt{2^{\tilde L_y/2 - 1}}} \sum_{\tau}  \mathbb{P}_{Q} \ket{\tau}_L \ket{\tau}_R,
\label{eqn:GSAppendix}
\end{align}
where $\ket{\tau}_L \in V_{0,L}^{\tilde{Q}}$ is a spin configuration on the $L$ spin chain. A similar statement holds for $\ket{\tau}_R$ on the $R$ spin chain. Let us call the operator $\tau_{\tilde{p},L}^{x} \tau_{\tilde{p},R}^{x}$ the `$x$-rung' operator and $\tau_{\tilde{p},L}^{z} \tau_{ \tilde{p}-1,L}^{z} \tau_{\tilde{p},R}^{z} \tau_{\tilde{p}-1,R}^{z}$ the `$z$-rung' operator.

For each topological sector $a$, we wish to calculate the corrections to the ground state of \eqn{eqn:HeffAppendix} in its $\mathcal{V}_{0,L}^{\tilde{Q}} \otimes \mathcal{V}_{0,R}^{\tilde{Q}}$ tensor product structure, generated by the small part of \eqn{eqn:fullHAppendix}. To leading order in $\epsilon$, the relevant terms in $H$ for each $\tilde{Q}$ sector that generate the changes in the two ground states of $\tilde{Q}$ are
\begin{align}
H_{\text{edge},L}^{\tilde{Q}} + H_{\text{edge},R}^{\tilde{Q}} + \epsilon (P_{0,L}^{\tilde{Q}} \otimes P_{0,L}^{\tilde{Q}}) V_{LR} (P_{0,L}^{\tilde{Q}} \otimes P_{0,L}^{\tilde{Q}}).
\end{align}
From the mapping to virtual spin operators acting on the two spin chains, we can rewrite the above as
\begin{align}
& f_{\tilde{Q},L}(\tau_{\tilde{p},L}^x, \tau_{\tilde{p},L}^z \tau_{\tilde{p}-1,L}^z ) + f_{\tilde{Q},R}(\tau_{\tilde{p},R}^x, \tau_{\tilde{p},R}^z \tau_{\tilde{p}-1,R}^z ) + \nonumber \\
&g_{\tilde{Q}}(\tau_{\tilde{p},L}^x, \tau_{\tilde{p},L}^z \tau_{\tilde{p}-1,L}^z , \tau_{\tilde{p},R}^x, \tau_{\tilde{p},R}^z \tau_{\tilde{p}-1,R}^z ),
\label{eqn:pertAppendix}
\end{align}
where $f_{\tilde{Q},L}$ and $f_{\tilde{Q},R}$ are as in \eqn{eqn:edgeHtildeQ} and $g_{\tilde{Q}}$ the function associated with $(P_{0,L}^{\tilde{Q}} \otimes P_{0,L}^{\tilde{Q}}) V_{LR} (P_{0,L}^{\tilde{Q}} \otimes P_{0,L}^{\tilde{Q}})$. There are also toroidal boundary conditions for both chains given by $\tilde{Q}_L$ (i.e. $\tau_{\tilde{0},L}^z = - \tau_{\tilde{L}_y/2,L}^z$) on the $L$ chain and similarly for the $R$ chain.

Since both the original, unperturbed Hamiltonian \eqn{eqn:HeffAppendix} and the perturbation \eqn{eqn:pertAppendix} are symmetric under $\mathbb{Z}_2 \times \mathbb{Z}_2$, generated by $\hat{Q}_L = \prod_{\tilde{p}} \tau_{\tilde{p},L}^x$ and $\hat{Q}_R = \prod_{\tilde{p}} \tau_{\tilde{p},R}^x$, it suffices to perform standard, non-degenerate wavefunction perturbation theory to the ground states \eqn{eqn:GSAppendix}, which have $Q_L = Q_R = Q$. We therefore need to solve for all the eigenstates of $H_{\text{eff}}$.

\vspace{5pt}
\noindent {\bf Eigenstates of $H_{\text{eff}}$.} $H_{\text{eff}}$  is exactly solvable because all terms in it commute. However there is one constraint: $\prod_{\tilde{p}=1}^{\tilde L_y/2} \tau_{ \tilde{p}, L}^{z} \tau_{ \tilde{p}-1,L}^{z} \tau_{\tilde{p},R}^{z} \tau_{\tilde{p}-1,R}^{z} = \mathds{1}$, and one term, $\hat{Q}_L$, not present in $H_{\text{eff}}$ that commutes with it. Thus, all eigenstates of $H_{\text{eff}}$ can be uniquely labeled by the eigenvalues of the set of commuting operators
\begin{align}
\left\{ \{ \tau_{ \tilde{p}, L}^x \tau_{\tilde{p},R}^x \}_{\tilde{p}=1}^{\tilde L_y/2}, \{ \tau_{ \tilde{p},L}^{z} \tau_{ \tilde{p}-1,L}^{z} \tau_{\tilde{p},R}^{z} \tau_{\tilde{p}-1,R}^{z} \}_{\tilde{p}=2}^{\tilde L_y/2},   \hat{Q}_L \right\}.
\end{align}
Note the choice of the $z$-rung operators from $\tilde{p} = 2$ to ${\tilde L_y/2}$ only. The ground states of $H_{\text{eff}}$  satisfy $\tau_{ \tilde{p}, L}^x \tau_{\tilde{p},R}^x = +1$ and $\tau_{ \tilde{p},L}^{z} \tau_{ \tilde{p}-1,L}^{z} \tau_{\tilde{p},R}^{z} \tau_{\tilde{p}-1,R}^{z} = +1$. There are two ground states given by \eqn{eqn:GSAppendix} with $Q = Q_L$.

Now consider the excited states of $H_{\text{eff}}$ that we will need in perturbation theory. Such states can be built up from the ground states by acting products of the following mutually commuting operators on them (they also commute with $\hat{Q}_L$):
\begin{align}
& s_{\tilde{p}}^z := \tau_{ \tilde{p},R}^z, \text{ for } \tilde{p} = \tilde{1}, \cdots, {\tilde L_y/2} \text { and } \nonumber \\
& s_{\tilde{p}}^x :=\prod_{\tilde{2} \leq \tilde{q} \leq \tilde {p}} \tau_{\tilde{p},L}^x \text{ for } \tilde{p} = \tilde{2}, \cdots, \tilde{L}_y/2,
\end{align}
to give $\prod \left( s_{\tilde{p}}^{z/x} \right) \ket{\tilde{Q}, Q}$. These $s$-operators violate the $x$-rung or $z$-rung $=+1$ conditions with an energy cost of $2g$ and $4g$ above the ground states respectively.

\vspace{5pt}
\noindent {\bf Ground state perturbation; calculation of $\Lambda'$.}
We are now ready to find the four perturbed ground states of the Wen-plaquette model on the infinite cylinder. Let us find the correction to the ground state $\ket{\tilde{Q},Q}$ for fixed $(\tilde{Q},Q)$.

Consider first the contribution from $H_{\text{edge},L}^{a} = \mathbb{P}_{Q,L} f_{\tilde{Q},L}(\tau_{\tilde{p},L}^x, \tau_{\tilde{p},L}^z \tau_{\tilde{p}-1,L}^z ) \mathbb{P}_{Q,L} $ first. $f_{\tilde{Q},L}$ contains products of $\tau_{L,\tilde{p}}^x$ and $\tau_{L, \tilde{p}}^z \tau_{L, \tilde{p}-1}^z$.  Let us consider a generic first order term in $f_{\tilde{Q},L}$ given by $\epsilon \big( c \prod \tau_{\tilde{p},L}^x \prod \tau_{\tilde{p},L}^z \tau_{\tilde{p}-1,L}^z \big)$, where $c$ is the coefficient of the term (necessarily making it hermitian).  From the first order {\it process} in perturbation theory, the correction to the ground state from this term is an order $\epsilon$ correction given by
\begin{align}
&\epsilon ~ c \sum_{\text{prod}}' \frac{ \bra{\tilde{Q},Q} \prod s^{z/x} \big( \prod \tau_{\tilde{p},L}^x \prod \tau_{\tilde{p},L}^z \tau_{\tilde{p}-1,L}^z \big) \ket{\tilde{Q},Q} }{\Delta(\prod s^{z/x} )} \times \nonumber \\
&\prod s^{z/x} \ket{\tilde{Q},Q}.
\label{eqn:corr}
\end{align}
An explanation is in order. The sum is over all possible products $\prod s^{z/x}$  which generate all possible excited states $\prod s^{z/x} \ket{\tilde{Q}, Q}$ (in the same topological sector), with the prime denoting the exclusion of the trivial product. $\Delta(\prod s^{z/x})$ is the energy gap between the ground state $\ket{Q}$ and  the excited state defined by the product, which is always negative: $\Delta < 0$. We have also made use of the fact that $\mathbb{P}_{Q,L} = 1$ on the particular ground state we are working with.

It is not hard to see that the only contributions arise if $\left( \prod s^{z/x} \right)  \left( \prod \tau_{\tilde{p},L}^x \prod \tau_{\tilde{p},L}^z \tau_{\tilde{p}-1,L}^z \right) = \mathds{1} \text{ or } \hat{Q}_L$. That is, the excited states cancel the excitations over the ground states from $H_{\text{edge},L}^{\tilde{Q}}$. In other words, we have
\begin{align}
&\prod s^{z/x} = \prod \tau_{\tilde{p},L}^x \prod \tau_{\tilde{p},L}^z \tau_{\tilde{p}-1,L}^z \text{ or} \nonumber \\
&\prod s^{z/x} = \hat{Q}_L \left( \prod \tau_{\tilde{p},L}^x \prod \tau_{\tilde{p},L}^z \tau_{\tilde{p}-1,L}^z \right).
\end{align}
For the first case, the matrix element is $1$, and so the excitation induced from $H_{\text{edge},L}^{\tilde{Q}}$ on the ground state is reproduced in the correction to the ground state, up to a negative rescaling that depends on the energy difference of this excited state with the ground state. For the second case, the matrix element is $Q$ (since $Q_L = Q$), and the correction to the ground state is 
\begin{align}
& \hat{Q}_L \left( \prod \tau_{\tilde{p},L}^x \prod \tau_{\tilde{p},L}^z \tau_{\tilde{p}-1,L}^z \right) \ket{\tilde{Q}, Q} \nonumber \\
= &\hat{Q}_L \left( \prod \tau_{\tilde{p},L}^x \prod \tau_{\tilde{p},L}^z \tau_{\tilde{p}-1,L}^z \right) Q \hat{Q}_L \ket{\tilde{Q}, Q} \nonumber \\
= &Q \left( \prod \tau_{\tilde{p},L}^x \prod \tau_{\tilde{p},L}^z \tau_{\tilde{p}-1,L}^z \right) \ket{\tilde{Q},Q},
\end{align}
where in the second equality we have commuted $\hat{Q}_L$ past the excitations. The $Q$ from the matrix element cancels the $Q$ from the correction to the ground statement, and so this correction once again reproduces the excitation induced from $H_{\text{edge},L}^{\tilde{Q}}$ on the ground state, up to a negative rescaling $1/\Delta(\prod s^{z/x})$. 

Interestingly, we therefore see that {\it each} term in the edge Hamiltonian $H_{\text{edge},L}^{\tilde{Q}}$ is reproduced as excitations to the ground state {\it acting on the $L$ spin chain}, albeit with a negative term-dependent rescaling $1/\Delta(\prod s^{z/x}) < 0$. That is, 

\noindent
\begin{align}
H_{\text{edge},L}^{a} &\xrightarrow{\text{induces corr.}} - \left( \bar{H}_{\text{edge,L}}^{\tilde{Q}} \right)_L \ket{\tilde{Q},Q} \nonumber \\
=& -\sum_{\tau', \tau} \left( \mathbb{P}_{Q} \bar{H}_{\text{edge,L}}^{\tilde{Q} } \mathbb{P}_Q \right)_{\tau', \tau} \ket{\tau'}_L \ket{\tau}_R,
\end{align}
where the the bar on $\bar{H}_{\text{edge},L}^{\tilde{Q}}$ signifies that we reproduce each term in $H_{\text{edge},L}^{\tilde{Q}}$ but with each term scaled by a positive rescaling: $1/|\Delta(\prod s^{z/x})|$. 

In the above, we have introduced notation using double subscripts (two $L$s). The $L$ in $\bar{H}_{\text{edge},L}^{\tilde{Q}}$ corresponds to the {\it form} of the Hamiltonian acting on the $L$ semi-infinite cylinder, while the $L$ of the parenthesis around it corresponds to operators acting on the $L$ spin-chain. Explicitly this means
\begin{align}
\left( \bar{H}_{\text{edge},\zeta}^{\tilde{Q}} \right)_\xi := \bar{H}_{\text{edge},\zeta}^{\tilde{Q}} ( \tau_{\tilde{p},\xi}^z \tau_{\tilde{p}-1,\xi}^z, \tau_{\tilde{p},\xi}^x),
\end{align}
where $\zeta, \xi \in \{ L, R \}$, and in the above case we have $\zeta = \xi = L$. We will use this notation below.

It will be helpful to provide examples of both cases to make our discussion concrete. Consider an example of the first case: a term $\tau_{\tilde{p},L}^x$ of $H_{\text{edge},L}^{\tilde{Q}}$ such that $\tilde{p} \neq \tilde{1}$. Then, there exists either a single operator ($s_{\tilde 2}^x$) or a product of two adjacent operators ($s_{\tilde{p}}^x s_{\tilde{p}-1}^x)$ which is the inverse of $\tau_{\tilde{p},L}^x$, i.e. itself. The energy difference $\Delta$ is $-4$ in this case. Consider next an example of the second case: $\tau_{L, \tilde{1}}^x$ in $H_{\text{edge},L}^{\tilde{Q}}$. One needs to multiply by $s_{\tilde L_y/2}^x$ to get $\hat{Q}_L$.  Now, $\bra{\tilde{Q}, Q}\hat{Q}_L \ket{\tilde{Q}, Q} s_{\tilde L_y/2}^x  \ket{\hat{Q}, Q}_L = Q_L^2 \left( s_{\tilde L_y/2}^x \hat{Q}_L \right) \ket{\tilde{Q},Q} = \tau_{ \tilde{1}, L}^x \ket{\tilde{Q}, Q}$. The excited state involved in the process also has an energy difference of $\Delta = -4$ from the ground state. Concluding, we have the result that {\it any}  $\tau_{\tilde{p},L}^x$ term that appears in $H_{\text{edge},L}^{\tilde{Q}}$ shows up also in the correction to the ground states:
\begin{align}
\tau_{\tilde{p},L}^x &\xrightarrow{\text{induces corr.}} -\frac{1}{4} \left( \tau_{\tilde{p},L}^x \right)_L \ket{\tilde{Q}, Q} \nonumber \\
= & -\frac{1}{4} \sum_{\tau', \tau} \left( \mathbb{P}_{Q} \tau_{\tilde{p},L}^x \mathbb{P}_Q \right)_{\tau', \tau} \ket{\tau'}_L \ket{\tau}_R
\end{align}

Next let us consider the corrections from $H_{\text{edge},R}^{\tilde{Q}}$. By making use of the fact the ground states satisfy the $x$-rung and $z$-rung operators $ = +1$, we can convert $H_{\text{edge},R}^{\tilde{Q}}$ acting on the {\it $R$ spin chain} to it acting on the {\it $L$ spin chain}:  $\prod \tau_{\tilde{p},R}^x \prod \tau_{\tilde{p},R}^z \tau_{\tilde{p}-1,R}^z \ket{\tilde{Q}, Q} = \prod \tau_{\tilde{p},L}^x \prod \tau_{\tilde{p},L}^z \tau_{\tilde{p}-1,L}^z \ket{\tilde{Q}, Q}$, and our above analysis holds. 

We therefore have
\noindent
\begin{align}
H_{\text{edge},R}^{a} &\xrightarrow{\text{induces corr.}} - \left( \bar{H}_{\text{edge},R}^{\tilde{Q}} \right)_L \ket{\tilde{Q},Q} \nonumber \\
=& -\sum_{\tau', \tau} \left( \mathbb{P}_{Q} \bar{H}_{\text{edge},R}^{\tilde{Q}} \mathbb{P}_Q \right)_{\tau', \tau} \ket{\tau'}_L \ket{\tau}_R,
\end{align}
where we remind the reader once again that it is the modified (term-dependent rescaled) $R$ edge Hamiltonian acting on the $L$ spin degrees of freedom of the ground state.

Lastly, consider the contribution from $\epsilon (P_{0,L}^{\tilde{Q}} \otimes P_{0,L}^{\tilde{Q}}) V_{LR} (P_{0,L}^{\tilde{Q}} \otimes P_{0,L}^{\tilde{Q}})$, which can be written as $g_{\tilde{Q}}(\tau_{\tilde{p},L}^x, \tau_{\tilde{p},L}^z \tau_{\tilde{p}-1,L}^z , \tau_{\tilde{p},R}^x, \tau_{\tilde{p},R}^z \tau_{\tilde{p}-1,R}^z )$ in the spin chain language.  Like above, if $g_{\tilde{Q}}$ acts on the ground state $\ket{\tilde{Q},Q}$, then we can convert terms that act on the $R$ spin chain to act on the $L$ spin chain, so that the overall contributions from $V_{LR}$ act only on the $L$ spin chain. For example, $\tau_{\tilde{p},L}^z \tau_{\tilde{p}-1,L}^z \tau_{\tilde{p},R}^x \ket{\tilde{Q},Q} =  \tau_{\tilde{p},L}^z \tau_{\tilde{p}-1,L}^z \tau_{\tilde{p},L}^x \ket{\tilde{Q},Q}$.

Thus,
\noindent
\begin{align}
V_{LR} &\xrightarrow{\text{induces corr.}} - \left( \bar{V}_{LR} \right)_L \ket{\tilde{Q},Q} \nonumber \\
=& -\sum_{\tau', \tau} \left( \mathbb{P}_{Q} \bar{V}_{LR} \mathbb{P}_Q \right)_{\tau', \tau} \ket{\tau'}_L \ket{\tau}_R,
\end{align}
where $\left( \bar{V}_{LR} \right)_L$ is a term-dependent rescaled, $R \to L$ operators swapped version of $V_{LR}$.

Therefore, we have calculated $\Lambda$ and hence $\Lambda'$:
\noindent
\begin{align}
\Lambda & =\mathbb{P}_Q ( \bar{H}_{\text{edge,L}}^{\tilde{Q}} + \bar{H}_{\text{edge,R}}^{\tilde{Q}} + \bar{V}_{LR}) \mathbb{P}_Q 
 \equiv \mathbb{P}_Q \Lambda' \mathbb{P}_Q.
\label{eqn:Lambda}
\end{align}
\vspace{5pt}

$ \left( \bar{H}_{\text{edge,L}}^{\tilde{Q}} + \bar{H}_{\text{edge,R}}^{\tilde{Q}} + \bar{V}_{LR} \right)_L $ differs from $\left(H_{\text{edge,L}}^{\tilde{Q}} + H_{\text{edge,R}}^{\tilde{Q}} \right)_L$ in two ways: the term dependent rescaling $|\Delta(\prod s^{z/x})|$, and also in a potentially arbitrary fashion from $\bar{V}_{LR}$. Thus, there is no reason to expect that the two spectra should match in the low-energy limit, leading to the conclusion that there is no edge-ES correspondence in general.

\section{Derivation of $H_{\text{edge},L}^a$ and $H_{\text{ent.}}^a$ for uniform magnetic fields as perturbations}
\label{sect:AppendixC}
In this appendix we calculate $H_{\text{edge},L}^a$ and $H_{\text{ent.}}^a$ for the case of perturbations being uniform single-site magnetic fields:
\begin{align}
\epsilon V = \epsilon \sum_{i} h_X X_i + h_Y Y_i + h_Z Z_i.
\end{align}
This can be written as $V = V_L + V_R$ where $V_L$ are the perturbations acting on the $L$ semi-infinite cylinder and $V_R$ are the perturbations acting on the $R$ semi-infinite cylinder.

\subsection{Calculation of $H_{\text{edge},L}^a$}
\label{sect:AppendixC_edge}
We calculate $H_{\text{edge},L}^a$ of the $L$ semi-infinite cylinder according to the Schrieffer-Wolff transformation, \eqn{eqn:HeffAppendixSW} and \eqn{eqn:EffHMethodAppendix}, for the perturbation $V_L$, to lowest non-trivial order.

The zeroth order term is identically $0$, since the unperturbed Wen-plaquette model on the  semi-infinite cylinder has a flat edge theory. Next, the first order ($\epsilon$) term of the edge Hamiltonian in each topological sector, $\epsilon P_{0,L}^a V_L P_0^a$, is also identically $0$ because single site magnetic fields cannot connect states in $\mathcal{V}_{0,L}^a$ to $\mathcal{V}_{0,L}^a$. Thus, we have to go to second order in $\epsilon$.

At this order, there can now be virtual processes that connect $\mathcal{V}_{0,L}^a$ to $\mathcal{V}_{0,L}^a$. They are mediated by excited states that are one unit of energy above the ground states. We can therefore simplify the the notation in \eqn{eqn:EffHMethodAppendix} for the case of the Wen-plaquette model, in each $\tilde{Q}$ sector:
\begin{align}
\bra{i,0,\tilde{Q}} H' \ket{j,0,\tilde{Q}} &=-\epsilon^2 \sum_{k} \bra{i,0,\tilde{Q}}V_L\ket{k,1} \bra{k, 1} V_L \ket{j, 0 , \tilde{Q}} \nonumber\\
& = -\epsilon^2 \bra{i,0,\tilde{Q}} V_L V_L \ket{j,0,\tilde{Q}},
\end{align}
for $\ket{i,0,\tilde{Q}} \in \mathcal{V}_{0,L}^{\tilde{Q}}$ and $\ket{k,1} \in \mathcal{V}_{1,L}$. 

Now, there are only two ways in which the above matrix element is non-zero. The first case is when a term in the first $V_L$ annihilates a term in the second $V_L$. However, this simply contributes to a diagonal matrix element whose value is the same for all states. In other words, such a process just renormalizes the energy of the system. We will hence disregard the contributions of this matrix element. The second case is when a term in the first $V_L$ combines with a term in the second $V_L$ to form a boundary operator $S_{p,L}$ or $S_{\tilde{p},L}$. This is now not a diagonal matrix element, and gives the lowest order non-trivial dynamics to the edge Hamiltonian. Dropping the label $\{0\}$ for notational convenience, the edge Hamiltonian to second order in $\epsilon^2$ simplifies to 
\begin{align}
&\bra{i,\tilde{Q}}H'\ket{j,\tilde{Q}} = -\epsilon^2  h_Z h_X  \sum_{l=1}^{L_y}  \times \nonumber \\
&\left( \bra{i,\tilde{Q}} Z_l X_{l+1}  \ket{j,\tilde{Q}} +  \bra{i,\tilde{Q}} X_{l+1} Z_{l}  \ket{j,\tilde{Q}} \right) \nonumber \\
& = -2 \epsilon^2  h_Z h_X  \sum_{l=1}^{L_y}  \bra{i,\tilde{Q}} Z_l X_{l+1}  \ket{j,\tilde{Q}} \nonumber \\
& = -2 \epsilon^2   h_Z h_X    \bra{i,\tilde{Q}} \hspace{-3pt} \left(  \sum_{p=1}^{L_y/2} \left( \scalebox{.55}{\EdgeRed{p} } \right)_L \hspace{-10pt} + \sum_{\tilde{p}=1 }^{\tilde L_y/2}  \left( \scalebox{.55}{\EdgeBlue{\tilde p} }\right)_L    \right)\hspace{-3pt}   \ket{j,\tilde{Q}}.
\end{align}
From the mapping, \eqn{eqn:Map}, we can read off the edge Hamiltonian in each $\tilde{Q}$ sector at order $\epsilon^2$:
\begin{align}
H_{\text{edge},L}^{\tilde{Q}} = -2 \epsilon^2 h_Z h_X \sum_{\tilde{p} = 1}^{\tilde L_y/2}   \left( \tau_{\tilde p,L}^z \tau_{\tilde{p}-1,L}^z +   \tau_{\tilde p,L}^x    \right),
\end{align}
with toroidal boundary conditions $\tau_{0,L} = \tilde{Q} \tau_{\tilde{L}_y/2,L}$. This is nothing but the critical periodic ($\tilde{Q} = +1$) or antiperiodic ($\tilde{Q} = -1$) $(1+1)$-d Ising model, which is $\mathbb{Z}_2$ symmetric and is clearly at the Kramers-Wannier self-dual point. 

The edge Hamiltonian in each topological sector $a$ is then found by projecting the Ising model down into the relevant $\mathbb{Z}_2 = Q_L$ sector:
\begin{align}
H_{\text{edge},L}^{a\simeq(\tilde{Q},Q)} = \mathbb{P}_{Q,L} H_{\text{edge},L}^{\tilde{Q}} \mathbb{P}_{Q,L},
\end{align}
where $\mathbb{P}_{Q} = (\mathds{1} + Q_L \hat{Q}_L)/2$, $\hat{Q}_L :=  \prod_{\tilde p} \tau_{\tilde p,L}^x$.

The edge Hamiltonian on the $R$ semi-infinite cylinder at $\epsilon^2$, $H_{\text{edge},R}^a$ is identical, except with the replacement $\vec{\tau}_{\tilde{p},L} $ with $\vec{\tau}_{\tilde{p},R}$.

\subsection{Calculation of $H_{\text{ent.}}^a$}
\label{sect:AppendixC_ent}
We calculate $H_{\text{ent.}}^a$ of the ground states of the Wen-plaquette model perturbed by uniform single-site magnetic fields to lowest order in $\epsilon$. We start with the calculation of $\Lambda'$ in \eqn{eqn:Lambda} of Appendix \ref{sect:AppendixB}. Firstly note that $V_{LR} = 0$, so $\tilde{V}_{LR} = 0$ as well. Secondly note that as discussed in Appenix \ref{sect:AppendixB}, each term in the both edge Hamiltonians is reproduced in $\Lambda'$, acting on the $L$ spin degrees of freedom, with a term-dependent rescaling that depends on the energy difference between the ground state of $H_{\text{eff}}$ in \eqn{eqn:HeffAppendix} and the excited states of $H_{\text{eff}}$ generated by the edge Hamiltonians. However, both edge Hamiltonians only contain the terms $\tau_{\tilde{p},L/R}^x$ and $\tau_{\tilde{p},L/R}^z \tau_{\tilde{p}-1,L/R}^z$, and the excited states generated by these terms acting on the ground states $\ket{\tilde{Q},Q}$ give an energy difference of $-4$ (they violate $2$ $z$-rung operators or $2$ $x$-rung operators respectively). Thus, in this case, the term-dependent rescaling becomes a single, overall rescaling, and we have the result
\begin{align}
\Lambda' = \frac{1}{4}\left( H_{\text{edge},L}^{\tilde{Q}} + H_{\text{edge},R}^{\tilde{Q}} \right)_L = \frac{1}{2}\left( H_{\text{edge},L}^{\tilde{Q}} \right)_L,
\label{eqn:LambdaPrime}
\end{align}
where in the second equality we made use of the fact that $ (H_{\text{edge},R}^{\tilde{Q}})_L = (H_{\text{edge},L}^{\tilde{Q}})_L$. 

At this point, it would be tempting to conclude from \eqn{eqn:Hent} and \eqn{eqn:pertES} that the entanglement Hamiltonian is then precisely proportional to the edge Hamiltonian in the lowest order in $\epsilon$. However, the identification $H_{\text{ent.}}^a \equiv 2 \mathbb{P}_Q \Lambda' \mathbb{P}_Q$ is only valid if $\Lambda'$ appears at order $\epsilon$. In the case we are considering here, it appears at order $\epsilon^2$, and so we cannot directly identify the entanglement Hamiltonian. 

To be consistent, we want to calculate all $\epsilon^2$ corrections to the reduced density matrix $\rho_L^a$. This implies that in \eqn{eqn:genericState}, we have to calculate $\Lambda$ to second order in $\epsilon$, and $\Theta$ and $\Xi$ to first order in $\epsilon$. Thus our calculation of $\Lambda'$ in \eqn{eqn:LambdaPrime} is not entirely correct as it is only a part of the $\epsilon^2$ correction. Also, it might be the case that $\Theta$ and $\Xi$ give rise to undesirable $\epsilon^2$ contributions in the dominant part of the reduced density matrix that modifies the entanglement spectrum from that of the spectrum of the edge Hamiltonians. However, we will show that in this case, (i) the other contribution to $\Lambda'$ is simply a constant, (ii) the contributions from $\Theta$ and $\Xi$ simply contribute shift the entanglement Hamiltonian. This calculation also explicitly shows how to perform calculations in our perturbative framework to order $\epsilon^2$, and by extension, to arbitrary order in $\epsilon$.

\noindent {\bf Result.} We first state the result. To order $\epsilon^2$,
\begin{align}
&\Lambda = \mathbb{P}_{Q,L} \Lambda' \mathbb{P}_{Q,L} = \mathbb{P}_{Q,L} \Bigg( \frac{1}{2} \left(   H_{\text{edge},L}^{\tilde{Q}} \right)_L  + \mathcal{O}(\epsilon^2)  \Bigg) \mathbb{P}_{Q,L} \nonumber \\
&\Theta = \epsilon \mathbb{P}_{Q,L} \otimes \vec{w}_1^\dagger \nonumber \\
&\Xi = \epsilon  \mathbb{P}_{Q,L} \otimes \vec{w}_2 \nonumber \\
& \Omega \sim \mathcal{O}(\epsilon^2) \Omega'
\label{eqn:ResultAppendix}
\end{align}
where $\vec{w}_1$ and $\vec{w}_2$ are (long) column vectors of unit strength denoting coefficients in front of a state $\ket{\tau}_L \ket{i,\alpha}_R$  $(\alpha \geq 1)$, and $ \ket{i,\alpha}_L \ket{\tau}_R$ respectively. The exact expression or length of $\vec{w}_1$ and $\vec{w}_2$ are unimportant here. $\Omega$ is a matrix that appears at $\epsilon^2$ as it requires at least a second order virtual process to create an excitation to the ground state outside of the space of two spin chains $\mathcal{V}_{0,L}^{\tilde{Q}} \otimes \mathcal{V}_{0,R}^{\tilde{Q}}$.

With this result, let us form the reduced density matrix $\rho_L^a = \Tr_R \ket{\tilde{Q},Q} \bra{\tilde{Q},Q}$. To order $\epsilon^2$, we have the unnormalized reduced density matrix
\begin{align}
&\rho_L^a =  \sum_{\tau',\tau} \bigg( \big(1+\mathcal{O}(\epsilon^2) + \vec{w}_1^\dagger \vec{w}_1 \big) \mathbb{P}_{Q,L}\nonumber \\
&  - \mathbb{P}_{Q,L} \left( H_{\text{edge},L}^{\tilde{Q}} \right)_L  \mathbb{P}_{Q,L} \bigg)_{\tau', \tau} \ket{\tau'}_L \bra{\tau}_L \nonumber \\
& + \sum_{\tau, i, \alpha \geq 1}  \left( \mathbb{P}_{Q,L}  \otimes \vec{w}_2 \ket{\tau}_L \bra{i,\alpha}_L + \text{h.c.} \right) \nonumber \\
& + \sum_{\stackrel{i,\alpha\geq 1}{j,\beta \geq1}} \cdots \mathcal{O}(\epsilon^2) \cdots \ket{i,\alpha}_L \bra{j, \beta}_L.
\end{align}
To extract the entanglement spectrum, we find the eigenvalues of $\rho_L^a$. This can be calculated in standard matrix perturbation theory, yielding the unnormalized eigenvalues
\begin{align}
&\text{eig.}(\rho_L^a) = \nonumber\\
&\bigg(1 + \mathcal{O}(\epsilon^2) + \vec{w}_1^\dagger \vec{w}_1 + \vec{w}_2^\dagger \vec{w}_2 \bigg) - \text{eig.}\bigg(  \mathbb{P}_{Q,L} \left( H_{\text{edge},L}^{\tilde{Q}} \right)_L  \mathbb{P}_{Q,L} \bigg).
\end{align}
Since $\vec{w}_1^\dagger \vec{w}_1$ and $\vec{w}_2^\dagger \vec{w}_2$ are just numbers of order $\epsilon^2$, we see that we can absorb the order $\epsilon^2$ constants in the first term of the above expression into a constant shift of the second term, which is nothing but the entanglement Hamiltonian. Thus, we have
\begin{align}
H_{\text{ent.}}^a = \mathbb{P}_{Q,L} \left( H_{\text{edge},L}^{\tilde{Q}} \right)_L  \mathbb{P}_{Q,L} + \text{const.},
\end{align}
at order $\epsilon^2$. Therefore, it is clear that in this case
\begin{align}
 H_{\text{ent.}}^a = H_{\text{edge},L}^a
\end{align}
up to shifting and rescaling, at order $\epsilon^2$ -  an edge-ES correspondence. The edge/ES Hamiltonians calculated in this case are the critical $(1+1)$-d Ising models, or the transverse field Ising model, \eqn{eqn:TFIM}, projected into the different charge sectors ($\mathbb{Z}_2$ labeled by $Q$ and toroidal boundary conditions labeled by $\tilde{Q}$). We have the following identification between states in the topological phase (left) and their edge/ES Hamiltonians (right):
\begin{align}
\ket{\mathbb{I}} &\leftrightarrow H_{\text{TFIM},+1}^{+1} \nonumber \\
\ket{e} &\leftrightarrow H_{\text{TFIM},-1}^{+1} \nonumber \\
\ket{m} &\leftrightarrow H_{\text{TFIM},+1}^{-1} \nonumber \\
\ket{\varepsilon} &\leftrightarrow H_{\text{TFIM},-1}^{-1},
\end{align}
where the label $\{ \mathbb{I}, m, \varepsilon = e \times m\}$ indicates that the states carry the corresponding anyonic flux.

\vspace{5pt}
\noindent {\bf Proof.} We present the proof of our assertion, \eqn{eqn:ResultAppendix}. First let us find the order $\epsilon$ corrections in $\Theta$ and $\Xi$. We identify the relevant terms in \eqn{eqn:fullHAppendix} that contribute. Let us concentrate on the contribution from perturbations on the right semi-infinite cylinder, $V_R$. The term that contributes is $[-S_R^a, \alpha P_{\alpha,R}^a]$, specifically, the order-$\epsilon$ term of $S_L^a$, which is given by \eqn{eqn:S1}. By virtue of the fact that $V_R$ is a sum of single-site magnetic fields which can only connect the subspace $\mathcal{V}_{0,R}^a$ to $\mathcal{V}_{1,R}^a$ through a single virtual process, we can further distill the relevant term:
\begin{align}
- \epsilon (P_{0,R}^a V_R P_{1,R}^a + \text{h. c.} )
\end{align}
is the term that contributes to $\Theta$ to order $\epsilon$. The correction induced is
\begin{align}
& \epsilon \sum_{ \tau,i} \bra{\tau}_L \bra{i, \alpha}_R  P_{1,R} V_R P_{0,R} \ket{\tilde{Q},Q} \times  \ket{\tau}_L \ket{i,\alpha}_R  \nonumber \\
& =  \epsilon \sum_{ \tau,i} \bra{\tau}_L \bra{i, \alpha}_R V_R \bigg( \sum_{\sigma} \mathbb{P}_{Q,L} \ket{\sigma}_L \ket{\sigma}_R \bigg) \times   \ket{\tau}_L \ket{i,\alpha}_R \nonumber \\
& = \epsilon \sum_{\tau,\sigma,i} \left( \mathbb{P}_{Q,L} \right)_{\tau,\sigma} \bra{i, \alpha}_R V_R \ket{\sigma}_R \times \ket{\tau}_L \ket{i,\alpha}_R.
\end{align}
where $\alpha = 1$. Now, $V_R = \sum_s (w_1^*)_s v^s_R$, where $v^s_R$ are single-site magnetic fields. For each $v^s_R$, $v^s_R$ acting on $\ket{\sigma}_R$ creates a unique excited state $v^s_R\ket{\sigma}_R \in \mathcal{V}_1$ which is unique - there is no other $v^{s'}_R$ such that $v_R \ket{\sigma}_R = v^{s'}_R \ket{\sigma}_R$. Thus, the label $s$ identifies a unique excited state $ \ket{\sigma_s^e}_R  \equiv v_R^s \ket{\sigma}_R$. Using this result, we can write the correction as
\begin{align}
\epsilon \sum_{\tau, s} \left( \mathbb{P}_{Q,L} \right)_{\tau,\sigma} (w_1^*)_s \ket{\tau}_L \ket{\sigma_s^e}_R,
\end{align}
from which we read off 
\begin{align}
\Theta = \epsilon \mathbb{P}_{Q,L} \otimes \vec{w}_1^\dagger,
\end{align}
as claimed. A similar analysis for the contributions from the perturbation $V_L$ on the $L$ semi-infinite cylinder will yield
\begin{align}
\Xi = \epsilon \mathbb{P}_{Q,L} \otimes \vec{w}_2.
\end{align}

Next, we show that $\Lambda$ has the asserted form. We have already accounted for the $H_{\text{edge},L}^{\tilde{Q}}$ term, as it appears from the edge Hamiltonians, and so we only have to account for the $\mathcal{O}(\epsilon^2)$ shift in $\Lambda'$ of \eqn{eqn:ResultAppendix}. 

This $\mathcal{O}(\epsilon^2)$ shift arises from the second order {\it process} in perturbation theory. This second order process corrects the state $\ket{n^{(0)}}$ as
\begin{align}
 & \Bigg( \sum_{\stackrel{k \neq n}{l \neq n}} \ket{k^{(0)}} \frac{\bra{k^{(0)}}V\ket{l^{(0)} } \bra{l^{(0)}}V\ket{k^{(0)} }  }{(E_n^{(0)} - E_k^{(0)} )(E_n^{(0)} - E_l^{(0)} ) } \nonumber \\
& - \sum_{k \neq n} \ket{k^{(0)}}  \frac{\bra{n^{(0)}}V\ket{n^{(0)} } \bra{k^{(0)}}V\ket{n^{(0)} }  }{(E_n^{(0)} - E_k^{(0)} )^2} \nonumber \\
& - \frac{1}{2} \ket{n^{(0)}} \sum_{k \neq n} \frac{\bra{n^{(0)}}V\ket{k^{(0)} } \bra{k^{(0)}}V\ket{n^{(0)} }  }{(E_n^{(0)} - E_k^{(0)} )^2} \Bigg),
\label{eqn:WFPert}
\end{align}
where $\ket{k^{(0)}}$ refers to eigenstates of the unperturbed Hamiltonian. As it is a process which involves two $V$s it gives rise to contributions of at least order $\epsilon^2$. In our case, the unperturbed Hamiltonian is the exact Wen-plaquette model $H_L + H_R + H_{LR}$, and $\ket{n^{(0)}}$ is the ground state of the exact Wen-plaquette model on the infinite cylinder in each topological sector, \eqn{eqn:GSAppendix}. Note that the second term evaluates to $0$ because $\bra{n^{(0)}}V\ket{n^{(0)} } = 0$.

Now, there are two contributions to $\mathcal{O}(\epsilon^2)$. The third term in \eqn{eqn:WFPert} simply rescales $\ket{n^{(0)}} = \ket{\tilde{Q},Q}$ - this gives one source of the shift $\mathcal{O}(\epsilon^2)$ in $\Lambda'$. The other source comes from the $[-S_L^a,[-S_L^a, \alpha P_{\alpha,L}^a]]$ term in \eqn{eqn:fullHAppendix} (and also the $R$ term), specifically with the first order term $S_1$ of $S_L^a$ and $S_R^a$. Expanding the two commutators and focusing on the relevant term on the $R$ cylinder gives us
\begin{align}
[-S_R^a,[-S_R^a, \alpha P_{\alpha,R}^a]] \sim \epsilon^2 P_{0,R}^a V_R P_{1,R}^a V_R P_{0,R}^a.
\end{align}
However making use of the fact that the each term in $V_R = \sum_s (w^*_1)_s v_R^a$ creates a unique excited state above any given state in $\mathcal{V}_{0,R}^a$, it must be that the above term is simply proportional (at order $\epsilon^2$) to $P_{0,R}^a$, i.e. 
\begin{align}
[-S_R^a,[-S_R^a, \alpha P_{\alpha,R}^a]] \sim \epsilon^2 P_{0,R}^a.
\end{align}
This then contributes $\sim \epsilon^2 \ket{\tilde{Q},Q}$ in perturbation theory as well. A similar story holds for the $L$ term. Thus, we have identified the sources of the $\mathcal{O}(\epsilon^2)$ shift in $\Lambda'$ of  \eqn{eqn:ResultAppendix}.

This concludes the proof of our claim.

\bibliography{refs}
%\bibliography
%\input{Z2_EdgeES.bbl}
%\bibliography{Z2_EdgeES}

\newpage

\end{document}